\documentclass[12pt]{article}
\usepackage{a4,amsmath,theorem,epsfig,times,euler,euscript,cite}
\renewcommand{\baselinestretch}{1.1}
\newcommand{\myTitle}[1]{\begin{center}{\bf\Huge #1}\\[5ex]\end{center}}
\newcommand{\myAuthor}[1]{\begin{center}{\Large #1}\\[2ex]\end{center}}
\newcommand{\myAffiliation}[1]{\\[1ex]{\it\large #1}}
\newcommand{\myEmail}[1]{\\[1ex]{\tt\large #1}}
\newcommand{\myDate}{\begin{center}{\large\today}\\[5ex]\end{center}}
\newcommand{\myAbstract}[1]{\begin{center}\renewcommand{\baselinestretch}{1}{\bf Abstract}\\[2ex]\parbox{0.8\linewidth}{\small\hspace{15pt} #1}\end{center}\vspace{\baselineskip}}
\newcommand{\myReport}[1]{\hspace{\fill} #1}
\newcommand{\myPreprint}[1]{}
\newcommand{\myKeywords}[1]{}
\newcommand{\myFigure}[1]{\begin{figure}\begin{center}#1\end{center}\end{figure}}

\newcommand{\fudgea}{\\[0.8ex]}
\newcommand{\fudgeb}{\\[-0.0ex]}
\newcommand{\Appendix}[1]{Appendix~\ref{#1}}
\newcommand{\Section}[1]{Section~\ref{#1}}

\newcommand{\Figure}[1]{Fig.\ref{#1}}
\newcommand{\Equation}[1]{Eq.(\ref{#1})}

\newcommand{\ie}{{\it i.e.}}

\newcommand{\dimeps}{\varepsilon}
\newcommand{\Ss}{S}
\newcommand{\Dim}{D}
\newcommand{\dirac}{\delta}
\newcommand{\causeps}{\epsilon}
\newcommand{\imag}{\mathrm{i}}

\renewcommand{\Im}{\mathrm{Im}}

\newcommand{\srac}[2]{{\textstyle\frac{#1}{#2}}}

\newcommand{\II}{I}
\newcommand{\mm}[1]{m_{#1}}
\newcommand{\pp}[1]{q_{#1}}
\newcommand{\qq}{l}
\newcommand{\nn}{n}

\newcommand{\Cc}{c}
\renewcommand{\Ss}{S}
\renewcommand{\AA}{A}
\newcommand{\vx}{\!\vec{\,x}}
\newcommand{\vy}{\!\vec{\,y}}
\newcommand{\vz}{\!\vec{\,z}}
\newcommand{\vb}{\!\vec{\,b}}
\newcommand{\vg}{\!\vec{\,\gamma}}
\newcommand{\vn}{\!\vec{\,\eta}}

\newcommand{\vc}{\!\vec{\,c}}
\newcommand{\vbeta}{\!\vec{\,\beta}}
\newcommand{\inp}{\!\cdot\!}
\newcommand{\Vol}[1]{\Omega_{#1}}
\newcommand{\intV}[2]{\int_{\Vol{#1}}\!\!\!\!\!d^{#1}{#2}\,}
\newcommand{\Ifour}{\II_{4}}
\newcommand{\Ithree}[1]{\II_{3}(#1)}
\newcommand{\Ccoeff}{c}
\newcommand{\Dfin}{\II_{4}^{{\mathrm{fin}}}}
\newcommand{\conv}{h}
\newcommand{\Afun}{A_{0}}
\newcommand{\Bfun}{B_{0}}
\newcommand{\Cfun}{C_{0}}
\newcommand{\Dfun}{D_{0}}
\newcommand{\den}{\mathrm{den}}

\begin{document}
%
%
\myReport{IFJPAN-IV-2010-7}
\myPreprint{}

\myTitle{%
{\sc OneLOop}: for the evaluation of\fudgea one-loop scalar functions
\footnotetext{%
This work was partially supported by RTN European Programme, MRTN-CT-2006-035505
 (HEPTOOLS, Tools and Precision Calculations for Physics Discoveries at  Colliders)
}
}

\myAuthor{%
A.\ van Hameren%
\myAffiliation{%
The H.\ Niewodnicza\'nski Institute of Nuclear Physics\fudgeb
Polisch Academy of Sciences\\
Radzikowskiego 152, 31-342 Cracow, Poland%
\myEmail{hameren@ifj.edu.pl}
}
}


\myDate

\myAbstract{%
{\sc OneLOop} is a program to evaluate the one-loop scalar 1-point,
2-point, 3-point and 4-point functions, for all kinematical
configurations relevant for collider-physics, and for any non-positive
imaginary parts of the internal squared masses. It deals with all UV
and IR divergences within dimensional regularization. Furthermore,
it provides routines to evaluate these functions using
straightforward numerical integration.
}

\myKeywords{}


\newpage
\tableofcontents

\section{Introduction}
%
The experiments at LHC for the study of elementary particles demand precise predictions of signals and potential backgrounds for new physics.
This implies that for many processes the hard part has to be evaluated at least at the next-to-leading (NLO) level in perturbation theory, and this includes processes with four or more particles in the final state.
There has been a considerable progress in this respect in recent years, resulting in full differential calculations for such processes
\cite{%
Bredenstein:2008zb,
Bredenstein:2009aj,
Bredenstein:2010rs,
Ellis:2009zw,
KeithEllis:2009bu,
Melnikov:2009wh,
Berger:2009zg,
Berger:2009ep,
Berger:2010vm,
Bevilacqua:2009zn,
Bevilacqua:2010ve,
Binoth:2009rv
}%
.
One of the bottlenecks that had to be resolved in order to achieve this was the calculation of the one-loop amplitude necessary to determine the virtual contribution, which in the mentioned publications involves at least 6-point functions.
Developments have reached the stage that by now public programs exist that, given the machinery to calculate tree-level amplitudes, can calculate one-loop amplitudes involving 6-point functions and beyond at a high level of automation \cite{Ossola:2007ax,Mastrolia:2010nb}.

Besides dealing with an increase in combinatorial complexity, moving from tree-level amplitudes to one-loop amplitudes demands the treatment of one-loop integrals.
The most successful approach has been to express the one-loop amplitudes as a linear combination of universal one-loop functions, dependent on the particular scattering process under consideration only through the value of kinematical input parameters.
The calculation of the one-loop functions is universal, while the calculation of their coefficients is not.
In the ``tensor-approach'', these functions are the one-loop tensor functions.
They involve a reduction scheme to express the tensor functions in terms of scalar functions, but this is universal.
In the ``unitarity-approach'' the universal functions are the one-loop scalar functions themselves.

As mentioned before, most approaches to calculate one-loop amplitudes demand the calculation of the one-loop scalar functions.
These may be infra-red (IR) divergent, and have to be dealt with within a regularization scheme.
For NLO calculations it then suffices to have available the scalar 1-point, 2-point, 3-point and 4-point functions.
Higher-point scalar functions can be expressed as linear combination of these.
The preferred regularization scheme for NLO calculations in QCD is dimensional regularization, and public libraries {\sc LoopTools}~\cite{Hahn:1998yk}, {\sc QCDLoop}~\cite{Ellis:2007qk} and {\sc OneLOop}~\cite{vanHameren:2009dr} deliver all necessary scalar function for LHC kinematics and real particle masses.
The first two rely on the public library {\sc FF} for finite scalar functions~\cite{vanOldenborgh:1989wn}.

Sooner or later, multi-particle NLO calculations will involve unstable particles.
The preferred method to perform such calculations in a consistent way is the ``complex-mass scheme'', which demands the evaluation of one-loop functions with complex internal masses~\cite{Denner:2005fg}.
{\sc LoopTools} provides those, including the most complicated finite 4-point function with 4 complex masses.
The latter has been presented in \cite{Nhung:2009pm}, which is the completion
and implementation of the work in~\cite{'tHooft:1978xw}.
Alternative formulas for the 4-point function with complex masses have been given recently in~\cite{Denner:2010tr}.

In this write-up, the extension of {\sc OneLOop} to deal with complex masses is presented.
Essentially the only missing ingredient was the 4-point function with an arbitrary number of complex masses.
It has now been implemented too, using the formulas from \cite{Nhung:2009pm}.
As mentioned before, {\sc OneLOop} provides all finite and IR-divergent scalar functions withing dimensional regularization.
It does not provide a complete treatment of exceptional phase space points corresponding to Landau singularities.
Non-singular exceptional phase space points are taken care of to a certain extend, by expressing formulas in terms of differences of two dilogarithms divided by the difference of their arguments, and using numerically stable implementations of these.
However, as argued in \cite{Denner:2010tr}, these exceptional cases do only rarely appear in practical calculations.

Since the analytic continuation and the implementation of the formulas for the complex routines is rather delicate, it is useful to have the means available to check their correctness numerically.
First of all, they have successfully been compared with existing implementations.
This includes a private code by A.~Denner~\cite{colilib}.
As an extra check, they have been compared with results obtained via straightforward numerical integration.
The methods applied to achieve this are presented in this write-up, and routines to evaluate all scalar functions involving at least two scales via numerical integration, thereby avoiding the evaluation of any dilogarithm, have been included in the {\sc OneLOop} library.
The actual integration is performed with the help of the {\sc Cuba}-library~\cite{Hahn:2004fe} and routines from the {\sc Quadpack} library~\cite{quadpack}.

{\sc OneLOop} can be obtained from
\begin{center}
{\tt http://helac-phegas.web.cern.ch/}
\end{center}

\section{Program description}
%
The main part of the package consists of 10 source files, written in {\tt FORTRAN}~{\tt 77} with common extensions, like \verb|enddo| statements and \verb|do while| statements.
Furthermore, long names are used, and underscores are used in names.
In particular, all names of subroutines, functions and common blocks begin with \verb|avh_olo_|.
It is in principle possible to compile the source files and link the object files directly to the main program the routines are supposed to be used in.
It may, however, be more practical to create a library with the makefile coming with the package, and link the library instead.
%

\subsection{The 1-point function\label{seca0}}
%
The one-loop scalar 1-point function is defined as
%
\begin{equation}
\raisebox{-13pt}{\epsfig{figure=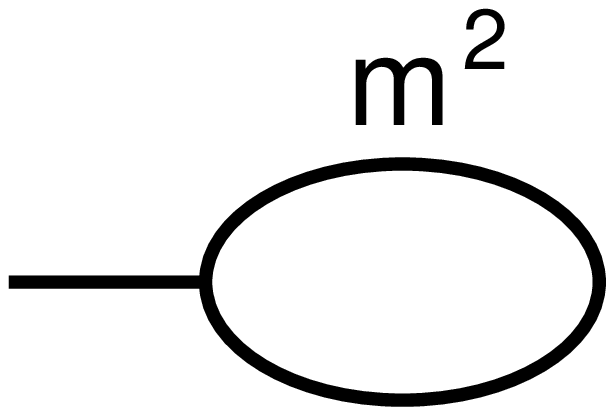,width=0.125\linewidth}}
\qquad
\Afun
=
\conv(\mu,\dimeps)\int
\frac{d^{4-2\dimeps}\qq}
     {\qq^2-\mm{}^2+\imag\causeps}
\label{defa0}
~,
\end{equation}
%
where the function $\conv$ of the renormalization scale $\mu$ and dimensional regularization parameter $\dimeps=2-\Dim/2$ follows the same convention as in \cite{Ellis:2007qk}, and is given by
%
\begin{equation}
\conv(\mu,\dimeps)
= \frac{\Gamma(1-2\dimeps)\,\mu^{2\dimeps}}
       {\Gamma^2(1-\dimeps)\Gamma(1+\dimeps)\,\imag\pi^{2-\dimeps}}
\label{defh}
~.
\end{equation}
%
The $\imag\causeps$ in the propagator denominators indicates the Feynman prescription for analytic continuation.
Squared momenta follow the Bjorken-Drell convention $\qq^2=\qq_0^2-\qq_1^2-\qq_2^2-\qq_3^2$.

The value of $\mu$ is by default equal to $1$.
This default can be changed with
\begin{verbatim}
subroutine avh_olo_mu_set( mu )
\end{verbatim}
The input \verb|mu| shall be real, and has the dimension of energy, not energy squared.

The 1-point function is evaluated with
\begin{verbatim}
subroutine avh_olo_a0c( rslt ,m )
\end{verbatim}
The input \verb|m| shall be complex, and is the squared mass $m^2$ in \Equation{defa0}.
It shall not have a positive imaginary part.
If it does, an error message is returned, and the sign of the imaginary part is considered to be the opposite.

The output \verb|rslt| shall be a complex array of shape \verb|(0:2)|, with \verb|rslt(0)| containing the $\dimeps^{0}$ coefficient and \verb|rslt(1)| containing the $\dimeps^{-1}$ coefficient of the Laurent expansion in $\dimeps$ of \Equation{defa0}.
The value of \verb|rslt(2)| is alway equal zero.
If \verb|m| is equal zero, then \verb|rslt(0)| and \verb|rslt(1)| are also zero.
%

\subsection{2-point functions}
%
The one-loop scalar 2-point function is defined as
%
\begin{equation}
\raisebox{-26pt}{\epsfig{figure=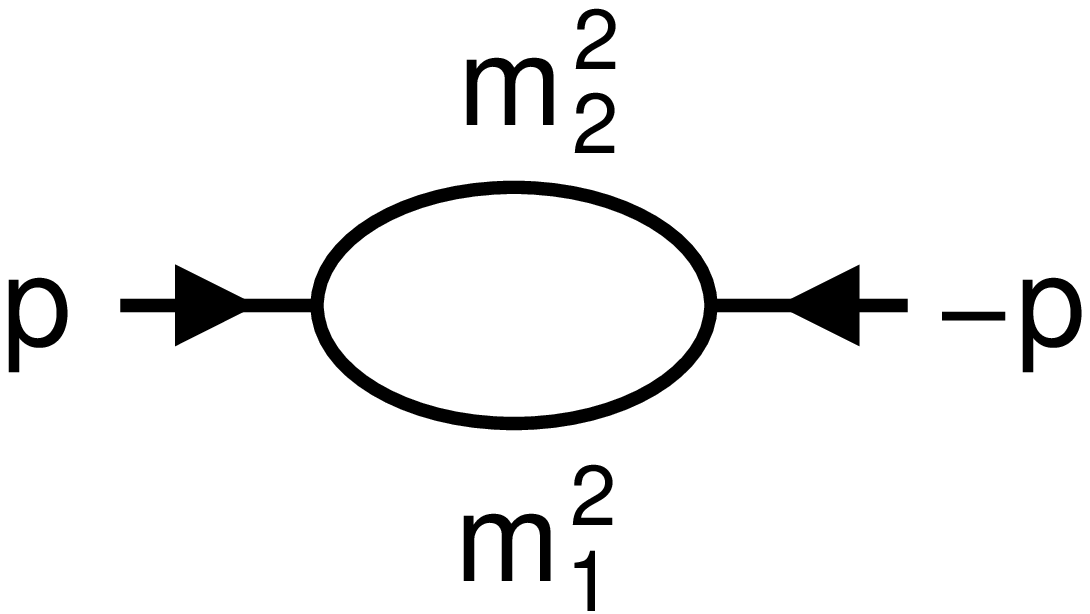,width=0.22\linewidth}}
\qquad
\Bfun
=
\conv(\mu,\dimeps)\int
\frac{d^{4-2\dimeps}\qq}
     {[\qq^2-\mm{1}^2+\imag\causeps][(\qq+p)^2-\mm{2}^2+\imag\causeps]}
\label{defb0}
~,
\end{equation}
%
where $\conv(\mu,\dimeps)$ is defined in \Equation{defh}.
The 2-point function is evaluated with
\begin{verbatim}
subroutine avh_olo_b0c( rslt ,p,m1,m2 )
\end{verbatim}
The input \verb|p,m1,m2| shall be complex, and is the square $p^2$ of the momentum and the squared masses $m_1^2$ and $m_2^2$ in \Equation{defb0} respectively.
The imaginary part of \verb|p| shall be identically zero.
If it is not, an error message is returned and the imaginary part is considered to be identically zero.
The squared masses shall not have a positive imaginary part.
If any of them does, an error message is returned, and the sign of the imaginary part is considered to be the opposite.

The output \verb|rslt| shall be a complex array of shape \verb|(0:2)|, with \verb|rslt(0)| containing the $\dimeps^{0}$ coefficient and \verb|rslt(1)| containing the $\dimeps^{-1}$ coefficient of the Laurent expansion in $\dimeps$ of \Equation{defb0}.
The value of \verb|rslt(2)| is alway equal zero.
If \verb|p| and \verb|m1| and \verb|m2| are equal zero, then \verb|rslt(0)| and \verb|rslt(1)| are also zero.

For the 2-point function, also the Passarino-Veltman functions $B_{1}$, $B_{00}$ and $B_{11}$ are available.
They are defined following
%
\begin{align}
\conv(\mu,\dimeps)\int
\frac{d^{4-2\dimeps}\qq\quad\qq^{\mu}}
     {[\qq^2-\mm{1}^2+\imag\causeps][(\qq+p)^2-\mm{2}^2+\imag\causeps]}
&=
p^{\mu}\,B_1
\\
\conv(\mu,\dimeps)\int
\frac{d^{4-2\dimeps}\qq\quad\qq^{\mu}\qq^{\nu}}
     {[\qq^2-\mm{1}^2+\imag\causeps][(\qq+p)^2-\mm{2}^2+\imag\causeps]}
&=
g^{\mu\nu}\,B_{00} +  p^{\mu}p^{\nu}\,B_{11}
~.
\end{align}
%
They are evaluated with
\begin{verbatim}
subroutine avh_olo_b11c( b11,b00,b1,b0 ,p,m1,m2 )
\end{verbatim}
The output \verb|b11,b00,b1,b0| respectively refers to $B_{11}$, $B_{00}$ and $B_{1}$ and $\Bfun$.
Each of them shall be a complex array of shape \verb|(0:2)| again, with the entries containing the coefficients of the Laurent expansion in $\dimeps$.

\subsection{The 3-point function}
%
The one-loop scalar 3-point function is defined as
%
\begin{equation}
\raisebox{-30pt}{\epsfig{figure=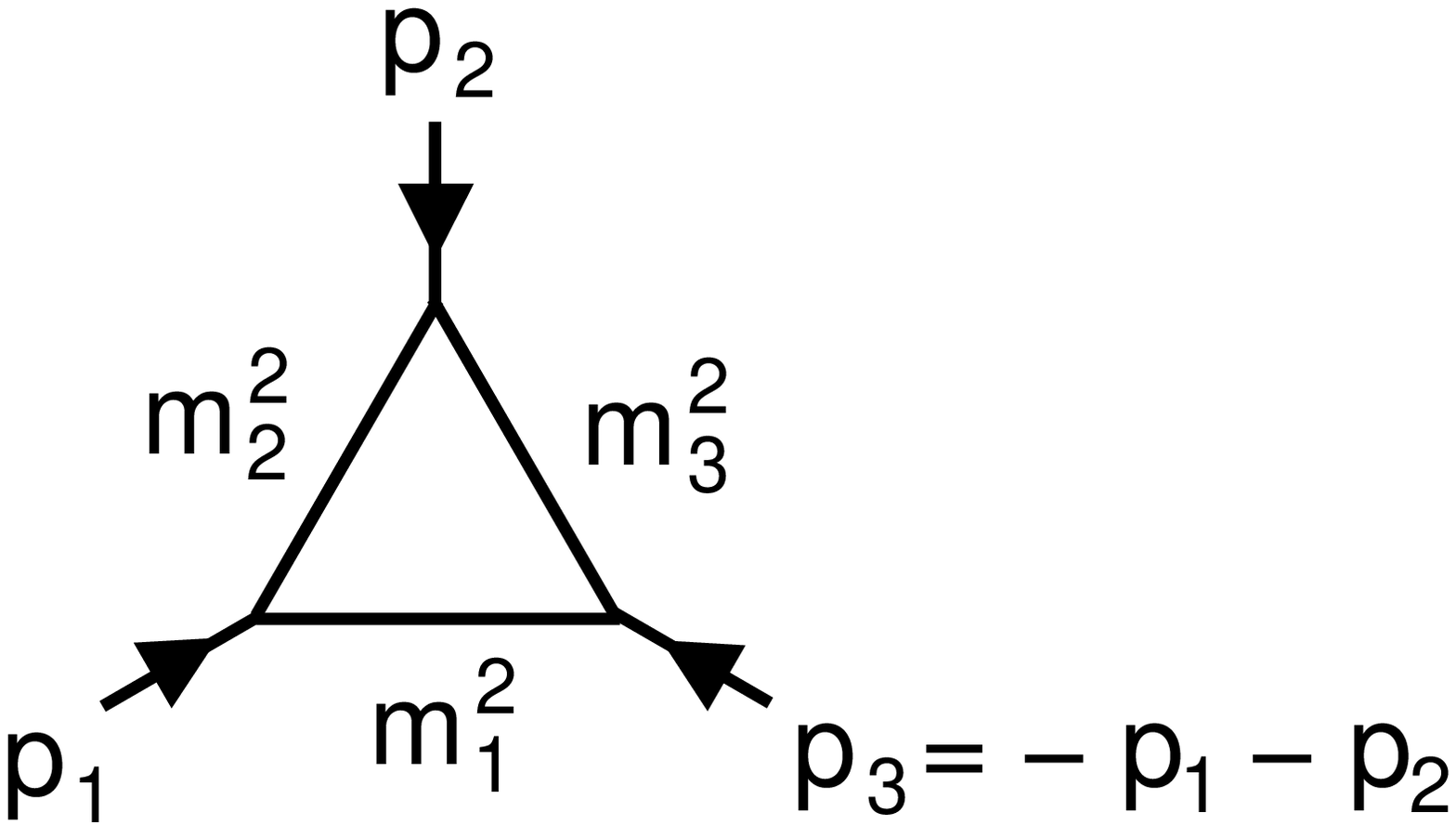,width=0.3\linewidth}}
\quad
\Cfun
=
\conv(\mu,\dimeps)\int
\frac{d^{4-2\dimeps}\qq}
     {\den_{1}(\qq)\,\den_{2}(\qq)\,\den_{3}(\qq)}
~,
\label{defc0a}
\end{equation}
%
where
%
\begin{align}
\den_{1}(\qq) &= \qq^2-\mm{1}^2+\imag\causeps           \nonumber\\
\den_{2}(\qq) &= (\qq+p_1)^2-\mm{2}^2+\imag\causeps     \label{defc0}         \\
\den_{3}(\qq) &= (\qq+p_1+p_2)^2-\mm{3}^2+\imag\causeps \nonumber
~,
\end{align}
%
and $\conv(\mu,\dimeps)$ is defined in \Equation{defh}.
The 3-point function is evaluated with
\begin{verbatim}
subroutine avh_olo_c0c( rslt ,p1,p2,p3 ,m1,m2,m3 )
\end{verbatim}
The input \verb|p1,p2,p3| shall be complex, and in terms of the momenta in \Equation{defc0} it is given by
%
\begin{equation}
p_1^2 \;\;,\;\; p_2^2 \;\;,\;\; p_3^2=(p_1+p_2)^2
~.
\end{equation}
%
The imaginary parts of these shall be identically zero.
If any of them is not, an error message is returned, and it is considered to be identically zero.

The input \verb|m1,m2,m3| shall also be complex, and in terms of the masses in \Equation{defc0} it is given by
%
\begin{equation}
m_1^2 \;\;,\;\; m_2^2 \;\;,\;\; m_3^2
~.
\end{equation}
%
The imaginary parts of these shall not be positive.
If any of them is, an error message is returned, and it will be considered to have the opposite sign.

The output \verb|rslt| shall be a complex array of shape \verb|(0:2)|, with \verb|rslt(0)| containing the $\dimeps^{0}$ coefficient, \verb|rslt(1)| containing the $\dimeps^{-1}$ coefficient, and \verb|rslt(2)| containing the $\dimeps^{-2}$ coefficient of the Laurent expansion in $\dimeps$ of \Equation{defc0a}.

By default, the values of \verb|rslt(1)| and \verb|rslt(2)| will be zero if the input does not {\em exactly\/} represent a configuration leading to an IR-divergent 3-point function. See \Section{secdiv} for more information.

\subsection{The 4-point function}
%
The one-loop scalar 4-point function is defined as
%
\begin{equation}
\raisebox{-37pt}{\epsfig{figure=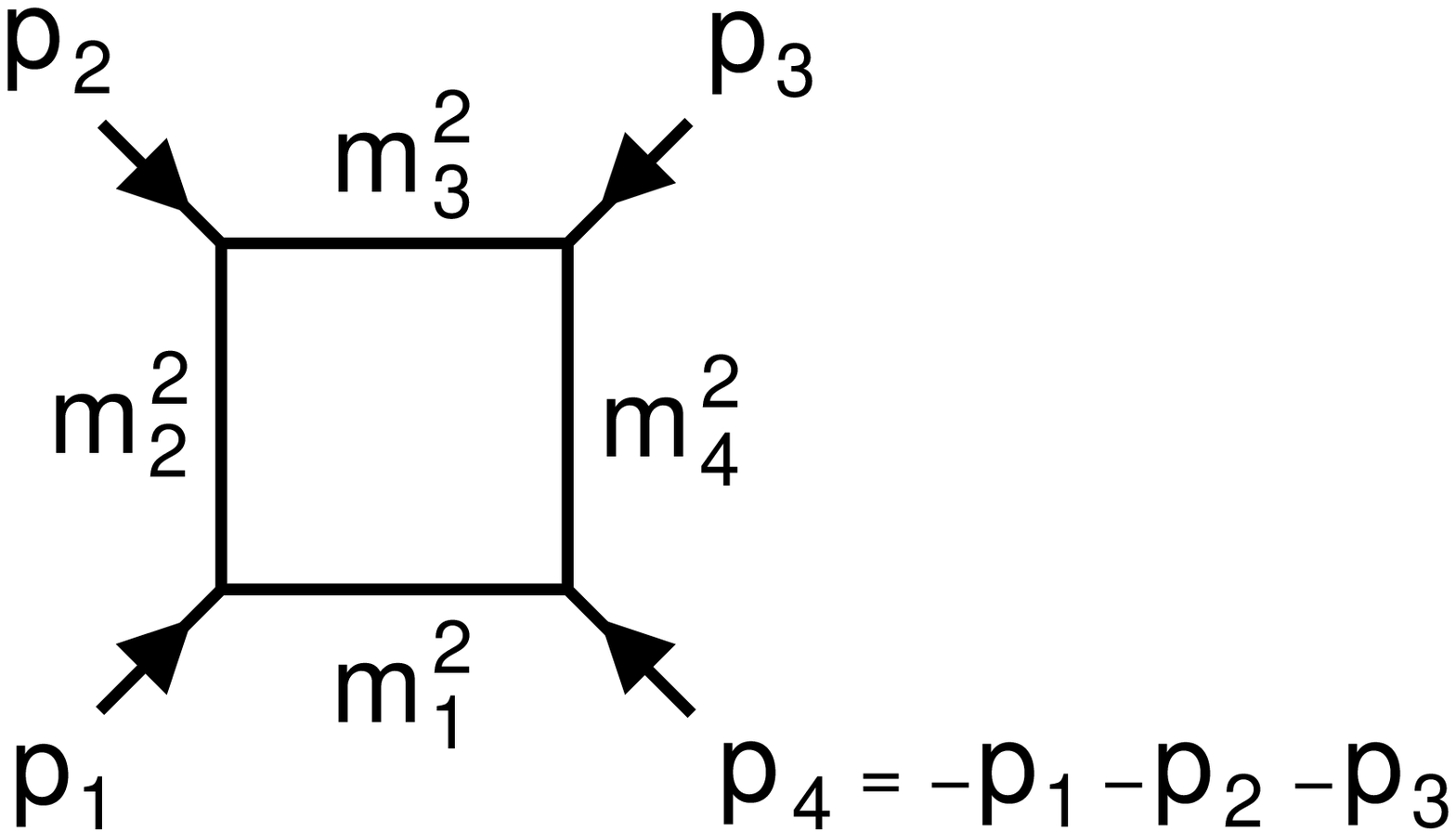,width=0.33\linewidth}}
\quad
\Dfun
=
\conv(\mu,\dimeps)\int
\frac{d^{4-2\dimeps}\qq}
     {\den_{1}(\qq)\,\den_{2}(\qq)\,\den_{3}(\qq)\,\den_{4}(\qq)}
~,
\label{defd0a}
\end{equation}
%
where
%
\begin{align}
\den_{1}(\qq) &= \qq^2-\mm{1}^2+\imag\causeps           \nonumber\\
\den_{2}(\qq) &= (\qq+p_1)^2-\mm{2}^2+\imag\causeps     \label{defd0}         \\
\den_{3}(\qq) &= (\qq+p_1+p_2)^2-\mm{3}^2+\imag\causeps \nonumber \\
\den_{4}(\qq) &= (\qq+p_1+p_2+p_3)^2-\mm{4}^2+\imag\causeps \nonumber
~,
\end{align}
%
and $\conv(\mu,\dimeps)$ is defined in \Equation{defh}.
The 4-point function is evaluated with
\begin{verbatim}
subroutine avh_olo_d0c( rslt ,p1,p2,p3,p4,p12,p23 ,m1,m2,m3,m4 )
\end{verbatim}
The input \verb|p1,p2,p3,p4,p12,p23| shall be complex, and in terms of the momenta in \Equation{defd0} it is given by
%
\begin{equation}
p_1^2 \;\;,\;\; p_2^2 \;\;,\;\; p_3^2 \;\;,\;\; p_4^2=(p_1+p_2+p_3)^2
\;\;,\;\; (p_1+p_2)^2 \;\;,\;\; (p_2+p_3)^2
~.
\end{equation}
%
The imaginary parts of these shall be identically zero.
If any of them is not, an error message is returned, and it is considered to be identically zero.

The input \verb|m1,m2,m3,m4| shall also be complex, and in terms of the masses in \Equation{defd0} it is given by
%
\begin{equation}
m_1^2 \;\;,\;\; m_2^2 \;\;,\;\; m_3^2 \;\;,\;\; m_4^2
~.
\end{equation}
%
The imaginary parts of these shall not be positive.
If any of them is, an error message is returned, and it will be considered to have the opposite sign.

The output \verb|rslt| shall be a complex array of shape \verb|(0:2)|, with \verb|rslt(0)| containing the $\dimeps^{0}$ coefficient, \verb|rslt(1)| containing the $\dimeps^{-1}$ coefficient, and \verb|rslt(2)| containing the $\dimeps^{-2}$ coefficient of the Laurent expansion in $\dimeps$ of \Equation{defd0a}.

By default, the values of \verb|rslt(1)| and \verb|rslt(2)| will be zero if the input does not {\em exactly\/} represent a configuration leading to an IR-divergent 4-point function. See \Section{secdiv} for more information.

\subsection{IR-divergent one-loop functions\label{secdiv}}
%
{\sc OneLOop} deals with all IR-divergent cases for the 3-point and the 4-point function within dimensional regularization.
IR-finite cases are defined as those for which the $\dimeps^{-1}$ coefficient and the $\dimeps^{-2}$ coefficient of the Laurent expansion in $\dimeps$ are identically zero.
IR-divergent cases are defined as those that are not IR-finite.
They are all listed in \cite{Ellis:2007qk}, and can also be found in \Appendix{AppBox} and \Appendix{AppTri}.

{\em By default, {\sc OneLOop} only returns IR-divergent cases if the input is exactly referring to a IR-divergent case.\/}
This does not mean that the input has to be given in a certain order.
Both
\begin{verbatim}
zero = dcmplx(0d0)
call avh_olo_c0c( rslt ,m2,p2,m3 ,zero,m2,m3 )
\end{verbatim}
and
\begin{verbatim}
zero = dcmplx(0d0)
call avh_olo_c0c( rslt ,m3,m2,p2 ,m3,zero,m2 )
\end{verbatim}
lead to a divergent case.
The default does, however, mean that the input has to be numerically exact.
For example, if an internal mass being equal to zero represents an IR-divergent case, then by default this divergent case is only returned if this mass is indeed identically zero on input.
Also if an external momentum squared being equal to an internal squared mass represents an IR-divergent case, this divergent case is only returned if these numbers are indeed identical on input.
{\em The default can be changed with}
\begin{verbatim}
subroutine avh_olo_onshell( thrs )
\end{verbatim}
The input \verb|thrs| shall be real, and is the threshold below which input is considered to be equal, or equal zero.
It carries the unit of energy squared, so if for example the absolute difference of a squared momentum and a squared mass is smaller than \verb|thrs|, they are considered to be equal.
Also, if a squared mass or squared momentum is absolutely smaller than \verb|thrs|, it is considered to be equal zero.
If the user decides to stick to the default, warnings are returned whenever the input is such that it is close to a divergent case.
The user is advised to take these warnings seriously.

The scale associated with the regularization of the IR divergences is the same as the renormalization scale $\mu$ of \Section{seca0}.
%

\subsection{Further useful routines}
%
All messages are by default sent to unit 6.
This default can be changed with
\begin{verbatim}
subroutine avh_olo_unit( unit )
\end{verbatim}
The input \verb|unit| shall be integer, and if it is larger than 0, messages will be sent to unit \verb|unit|.
If it is not larger than 0, no messages will be printed at all.

{\sc OneLOop} can be ordered to print all input and output appearing during a run with
\begin{verbatim}
subroutine avh_olo_printall( unit )
\end{verbatim}
The integer input \verb|unit| is the unit to which all this information is sent.
Again, if \verb|unit| is not larger than 0, nothing will be printed.

\subsection{Real routines}
%
Some users may want to stick to real input.
For all routines to evaluate one-loop functions, there are wrappers prepared that ask for strictly real input.
For historical reasons, the routine names end on \verb|m| rather than on \verb|r|, and are given by
\begin{verbatim}
subroutine avh_olo_a0m( rslt ,m )
subroutine avh_olo_b0m( rslt ,p,m1,m2 )
subroutine avh_olo_b11m( b11,b00,b1,b0 ,p,m1,m2 )
subroutine avh_olo_c0m( rslt ,p1,p2,p3 ,m1,m2,m3 )
subroutine avh_olo_d0m( rslt ,p1,p2,p3,p4,p12,p23 ,m1,m2,m3,m4 )
\end{verbatim}
%

\subsection{Routines using numerical integration}
%
As mentioned in the introduction, with the {\sc OneLOop} package also come routines to evaluate one-loop functions using numerical integration.
These have been included as means of a cross-check.
The main purpose is to check the correctness of analytic continuations, and the emphasis has been put on straightforwardness rather than sophistication in their implementation.
In particular the routines for the 4-point function can be slow, and will give only the first few decimals correctly.
Since errors in analytic continuation lead to rather drastic differences, not just in the last few decimals, this is in general sufficient.

The routines come as an independent library, which from user point of view essentially is a copy of the main {\sc OneLOop} library.
It has all routines, except those for the 2-point Passarino-Veltman functions, with the only difference that the routine names start with \verb|avh_oni| instead of \verb|avh_olo|.
It has to be compiled separately, and needs access to the \verb|cure| routine of the {\sc Cuba} library ~\cite{Hahn:2004fe}.
It provides two extra routines, namely
\begin{verbatim}
subroutine avh_oni_maxeval_set( maxeval )
subroutine avh_oni_unitcuba_set( unit )
\end{verbatim}
The integer input \verb|maxeval| of the first routine changes the maximal number of integrand evaluations \verb|cure| is ordered to perform, which by default is $10^6$.
The integer input \verb|unit| of the second routine changes the unit to which messages related to the numerical integration are sent, which is $6$ by default.
If this number is not larger than 0, no messages will be printed.

\subsection{Quadruple precision}
%
Some compilers provide the option to compile at a higher precision level, such that for example all 16-byte complex variables become 32-byte variables.
All series expansions inside {\sc OneLOop} have been prepared to also deal with this precision level.
During its first execution, {\sc OneLOop} will determine the precision level at which it has been compiled, and from then on use the relevant expansions.
%

\section{One-loop scalar functions via numerical integration}
%
In this section, integral representations for the one-loop scalar functions suitable for numerical integration are derived.
Since the aim is to have an alternative numerical implementation of the scalar functions in order to check the correctness of the analytic continuations in the ``hard-wired'' implementation, the main guideline in the derivation is that it should be straightforward.
Methods to smoothen the integrand in order to speed up numerical integration, as for example used in \cite{Passarino:2001wv,Ferroglia:2002mz}, are not applied.
Other approaches to numerical integration of loop integrals can be found in \cite{Nagy:2003qn,Nagy:2006xy,Gong:2008ww,Binoth:2005ff,Kurihara:2005ja,Anastasiou:2007qb}.

The main problem in the numerical integration of loop integrals is the existence of singularities in the integrand.
These can be categorized into two groups: those that can be dealt with via deformation of the integration contour, and those that cannot.
The former will be dealt with{\ldots}via contour deformation.
Among the latter are the so-called mass-singularities, or infra-red (IR) singularities.
They cause the integral to be divergent, and are typically inherent to the particular scattering process under consideration.
They ask for a regularization scheme, and are dealt with within dimensional regularization in this section.
Other singularities that cannot be dealt with via contour deformation typically appear in isolated points in phase space.
A complete treatment of those, like for example in \cite{Ferroglia:2002mz}, is beyond the scope of this work.

In this section, we write the one-loop $\nn$-point scalar integral within dimensional regularization in the following convention
%
\begin{equation}
\II_{\nn}
=
\frac{(-)^{\nn}}{\Gamma(\nn-2+\dimeps)\,\imag\pi^{2-\dimeps}}
\int\frac{d^{4-2\dimeps}\qq}
         {\prod_{j=1}^{\nn}[(\qq+\pp{j})^2-\mm{j}^2+\imag\causeps]}
\label{defI00}
~.
\end{equation}
%
The denominator momenta $\pp{j}$ are related to the external momenta $p_{j}$ via $\pp{1}=0$ and
%
\begin{equation}
\pp{j} = \sum_{k=1}^{j-1}p_{k}
~.
\end{equation}
%
The reason for the choice of the factor in front of the integral is that it leads to a simple expression for the scalar function in terms of the so-called Feynman parameters, namely
%
\begin{equation}
\II_{\nn}
=
\intV{\nn}{x}
\frac{\dirac\big(1-\sum_{j=1}^{\nn}x_{j}\big)}
     {(\vx\inp\Ss\vx - \imag\causeps)^{\nn-2+\dimeps}}
\label{Eq126}\end{equation}
%
where $\Ss$ is the linear operator represented by the symmetric $\nn\times\nn$ matrix with elements
%
\begin{equation}
\Ss_{j,k} = \frac{\mm{j}^2+\mm{k}^2-(\pp{j}-\pp{k})^2}{2}
~.
\label{Eq173}
\end{equation}
%
The integration region is such that
all variables $x_{j}$ run from $0$ to $\infty$,
%
\begin{equation}
\Vol{\nn} = [0,\infty)^{\nn}
\label{Eq184}
~,
\end{equation}
%
however the Dirac delta-function in \Equation{Eq126} confines the integration region eventually to the $\nn$-simplex.
One could proceed by direct elimination of one of the integration variables with the help of the delta-function.
We shall take a slightly different strategy, and scale the variables $x_{1}$ to $x_{\nn-1}$ by $x_{\nn}$, which we call $y$, to find
%
\begin{equation}
\II_{\nn}
=
\int_{0}^{\infty}dy\,y^{\nn-1}\intV{\nn-1}{x}
\frac{\dirac\big(1-y\big(1+\sum_{j=1}^{\nn-1}x_{j}\big)\big)}
     {y^{2\nn-4+2\dimeps}\,(\vx\inp\AA\vx+2\vb\inp\vx+\Cc)^{\nn-2+\dimeps}}
~.
\end{equation}
%
Here, we denote
%
\begin{equation}
\AA
=
\left( \begin{array}{cccc}
\Ss_{1,1}     & \Ss_{1,2}     & \cdots & \Ss_{1,\nn-1}    \\
\Ss_{2,1}     & \Ss_{2,2}     & \cdots & \Ss_{2,\nn-1}    \\
\vdots        & \vdots        & \ddots & \vdots           \\
\Ss_{\nn-1,1} & \Ss_{\nn-1,2} & \cdots & \Ss_{\nn-1,\nn-1}
\end{array} \right)
\quad,\quad
\vb
=
\left( \begin{array}{c}
\Ss_{\nn,1}    \\
\Ss_{\nn,2}    \\
\vdots         \\
\Ss_{\nn,\nn-1}
\end{array} \right)
\quad,\quad
\Cc
=
\Ss_{\nn\nn}-\imag\causeps
~.
\end{equation}
%
We used the fact that we are allowed to replace $\causeps$ by $y^2\causeps$ since $y^2$ is positive.
Elimination of the $y$-integral now gives
%
\begin{equation}
\II_{\nn}
=
\intV{\nn-1}{x}
\frac{\big(1+\sum_{j=1}^{\nn-1}x_{j}\big)^{\nn-4+2\dimeps}}
     {(\vx\inp\AA\vx+2\vb\inp\vx+\Cc)^{\nn-2+\dimeps}}
~.
\label{Eq217}
\end{equation}
%
The integration region is not finite anymore.
One advantage of expressing the one-loop integral as \Equation{Eq217} is that one does not have to bother about the end-point at infinity if one wants to explicitly deform the integration contour in the complex plane in order to avoid poles as prescribed by $\imag\causeps$.
It does not matter towards which direction in the complex plane infinity is approached, since the integrand does not have a singularity at infinity.
Another advantage is that a proliferation of terms is prevented when one wants to apply techniques involving integration by parts, since integrands will be $0$ at infinity.
%
\subsection{Contour deformation}
%
The $\imag\causeps$-prescription dictates that the contour deformation has to be such that the imaginary part of the polynomial in the denominator of the integrand in \Equation{Eq217} is negative when the real part vanishes.
Furthermore, the deformation has to be such that the end-points at $x_j=0$ (the begin-points) stay fixed.
IR-singularities, which cannot be dealt with by contour deformation, may show up as vanishing real parts at $x_j=0$.

Since we want to allow for squared masses with non-zero, but negative, imaginary parts, the matrix $\AA$ and the vector $\vb$ may have imaginary parts.
We re-define our notation, and write
%
\begin{equation}
\AA \leftarrow \AA - \imag\Gamma
\quad,\quad
\vb \leftarrow \vb - \imag\vg
~,
\end{equation}
%
now with $\AA$ and $\vb$ real.
The matrix $\Gamma$ and the vector $\vg$ have only non-negative components.
Our deformation is very similar to the one in \cite{Binoth:2005ff}.
We introduce a strictly increasing function $\tau$ such that $\tau(0)=0$ and $\tau(\infty)$ is finite, for example $\tau(x)=\tau_{\infty}x/(1+x)$ with $\tau_{\infty}>0$.
The deformation is given by
%
\begin{equation}
\vx \mapsto \vz(\vx) = \vx - \imag\vy(\vx)
~,
\end{equation}
%
with
%
\begin{equation}
y_j(\vx)
=
\tau(x_j)(\AA\vx+\vb)_j
~,
\label{Eq347}
\end{equation}
%
and where $(\AA\vx+\vb)_j$ denotes the $j$-th component of the vector $\AA\vx+\vb$.
The Jacobian matrix coming with the deformation is given by
%
\begin{equation}
\frac{\partial{}z_j}{\partial{}x_k}
=
\left[1-\imag\tau'(x_j)(\AA\vx+\vb)_j\right]\delta_{j,k}
-\imag\tau(x_j)\AA_{j,k}
~.
\end{equation}
%
The imaginary part of the polynomial in the denominator of the integrand in \Equation{Eq217} satisfies
%
\begin{equation}
-\Im\left\{\vz\inp(\AA-\imag\Gamma)\vz+2(\vb-\imag\vg)\inp\vz+\Cc\right\}
=
2\vy\inp(\AA\vx+\vb) - \vy\inp\Gamma\vy
+\vx\inp\Gamma\vx + 2\vg\inp\vx - \Im\{c\}
~.
\label{Eq360}
\end{equation}
%
The last three terms are safe, in the sense that they never become negative.
The first term never becomes negative if $\vy$ is chosen as in \Equation{Eq347} (notice that the function $\tau$ acts as a deformation of the inner product, keeping it however positive definite).
We have to worry a bit about the second term.
The matrix $\Gamma$ has elements
%
\begin{equation}
\Gamma_{j,k} = \eta_j + \eta_k
\quad,\quad
\eta_j = -\srac{1}{2}\Im\{m_j^2\} \geq0
~,
\end{equation}
%
so
%
\begin{equation}
\vy\inp\Gamma\vy
=
2\left(\sum_{j=1}^{\nn-1}y_j\right)\left(\sum_{j=1}^{\nn-1}\eta_j{}y_j\right)
~.
\end{equation}
%
Applying the Schwartz inequality to both sums, we find
%
\begin{equation}
|\vy\inp\Gamma\vy|
\leq
2\sqrt{\nn-1}\,\|\vn\|\,\|\vy\|^2
\quad,\quad
\|\vn\|^2 = \sum_{j=1}^{\nn-1}\eta_j^2
~.
\end{equation}
%
Now we have for the first two terms on the r.h.s.\ of \Equation{Eq360}
%
\begin{equation}
2\vy\inp(\AA\vx+\vb) - \vy\inp\Gamma\vy
\geq
2\sum_{j=1}^{\nn-1}\left( \frac{y_j^2}{\tau(x_j)} - \sqrt{\nn-1}\,\|\vn\|\,y_j^2 \right)
~,
\end{equation}
%
and we conclude that we are safe of we choose $\tau$ such that
%
\begin{equation}
\tau(\infty) \leq \frac{1}{\sqrt{\nn-1}\,\|\vn\|}
=
\left((\nn-1)\sum_{j=1}^{\nn-1}\Im\{m_j^2\}^2\right)^{-1/2}
~.
\end{equation}
%

Many methods for numerical integration ask for integrands living on the unit hypercube.
For our application, this can conveniently be achieved by preceding the contour deformation with the mapping
%
\begin{equation}
[0,1] \mapsto [0,\infty)
\quad,\quad
\rho_j \mapsto x_j = 1/\rho_j - 1
~,
\end{equation}
%
leading to a Jacobian weight factor $\rho_j^{-2}=(1+x_j)^2$ for each integration variable $\rho_j$ in the integrand.
%

\subsection{IR-divergent scalar integrals}
%
For IR-finite integrals, the integrand in \Equation{Eq217} can be expanded in $\dimeps$ and integrated term by term.
In particular, $\dimeps$ can be put to $0$ to calculate the $\dimeps^0$-term.
For IR-divergent integrals, this is not directly possible.
In this work, we are only interested in scalar integrals up to $\nn=4$, \ie\ box integrals and triangle integrals.
IR-divergent bubble integrals vanish within dimensional regularization.

Regarding the box integrals, we use the fact that the IR-divergent structure of one-loop integrals with more than $3$ external lines can be expressed in terms of triangle integrals.
This is completely worked out in \cite{Dittmaier:2003bc}.
Formulated for our application, that work explains how to determine the coefficients $\Ccoeff_j$ for every IR-divergent box integral such that it can be written as
%
\begin{equation}
\Ifour
=
\sum_{j=1}^{4}\Ccoeff_j\,\Ithree{j}
-\Dfin
\label{Eq489}
\end{equation}
%
where $\Ithree{j}$ is the triangle integral obtained from the box integral by removing propagator denominator $j$, and where $\Dfin$ is IR-finite.%
\footnote{Within the convention of \Equation{defI00}, $\Ithree{j}$ actually is a triangle integral divided by $-(1+\dimeps)$.}
Putting the terms under one integral and with a common denominator, we can write%
\begin{equation}
\Dfin
=
\sum_{j=1}^{4}\Ccoeff_j\,\Ithree{j} - \Ifour
=
\frac{1}{\Gamma(2+\dimeps)\,\imag\pi^{2-\dimeps}}
\int{}d^{4-2\dimeps}\qq\,
\frac{\alpha{}l^2 + 2v^{\mu}l_{\mu} + \gamma}
     {\prod_{j=1}^{4}[(\qq+\pp{j})^2-\mm{j}^2+\imag\causeps]}
~,
\label{Eq516}
\end{equation}
%
with
%
\begin{equation}
\alpha = \sum_{j=1}^{4}\Ccoeff_j
\quad,\quad
v^{\mu} = \sum_{j=1}^{4}\Ccoeff_j\pp{j}^{\mu}
\quad,\quad
\gamma = \sum_{j=1}^{4}\Ccoeff_j(\pp{j}^2-\mm{j}^2) - 1
~.
\end{equation}
%
The integral in \Equation{Eq516} is IR-finite, and can be integrated numerically.
The appearance of the term $\gamma$ in the numerator seems counter-intu\"\i{}tive in this respect, and in fact the coefficients $\Ccoeff_j$ appear to be such that it vanishes for every IR-divergent box integral.
In terms of Feynman parameters, $\II_{4}^{\mathrm{fin}}$ is given by
%
\begin{equation}
\II_{4}^{\mathrm{fin}}
=
\intV{4}{x}\dirac\bigg(1-\sum_{j=1}^{4}x_{j}\bigg)
\frac{\alpha\big( \vx\inp{}Q\vx - \srac{2}{1+\dimeps}\,\vx\inp\Ss\vx \big)
      - \vbeta\inp\vx + \gamma}
     {(\vx\inp\Ss\vx - \imag\causeps)^{2+\dimeps}}
~,
\label{Eq542}
\end{equation}
%
where the matrix $Q$ and the vector $\vbeta$ have components
%
\begin{equation}
Q_{j,k} = \pp{j,\mu}\pp{k}^{\mu}
\quad,\quad
\beta_j = 2v_{\mu}\pp{j}^{\mu}
~.
\end{equation}
%

To summarize, the IR-divergent box integrals are calculated following \Equation{Eq489}, where $\Dfin$ is finite and can be integrated numerically directly.
The coefficients $\Ccoeff_j$ for the triangle integrals can be determined as explained in \cite{Dittmaier:2003bc}, and are explicitly given in \cite{Denner:2010tr}.
They can also be found in \Appendix{AppBox}, together with the coefficients $\alpha$ and $\beta_j$.
For the IR-divergent triangle integrals themselves integral representations are derived in \Appendix{AppTri} which can be performed numerically.
%

\section{Summary}
%
The program {\sc OneLOop} for the evaluation of one-loop scalar functions up to and including 4-point functions was presented.
It deals with all kinematical configurations relevant for collider-physics, and any non-positive imaginary parts for the internal squared masses, and it deals with all UV and IR divergences within dimensional regularization.
Furthermore, it provides routines to evaluate all these functions using straightforward numerical integration, and the methods applied to achieve this were presented.
The program can be obtained from
\begin{center}
{\tt http://helac-phegas.web.cern.ch/}
\end{center}


%
%
\begin{appendix}

\section{Coefficients for IR-divergent box integrals\label{AppBox}}
%
In this appendix, the coefficients $\Ccoeff_j$ for \Equation{Eq489} and $\alpha$, $\beta_j$ for \Equation{Eq542} are given.
The coefficient $\gamma$ in the latter equation vanishes for all cases.
The coefficients are expressed in terms of the external momenta defined as in \Equation{defd0a}, and we denote
%
\begin{equation}
s_{12} = (p_1+p_2)^2
\quad,\quad
s_{23} = (p_2+p_3)^2
~.
\end{equation}
For completeness, we also give the matrix $\Ss$ in \Equation{Eq542}
%
\begin{equation}
\frac{1}{2}
\left( \begin{array}{cccc}
2m_1^2                   &  m_1^2+m_2^2-p_1^2        &  m_1^2+m_3^2-s_{12}  &  m_1^2+m_4^2-p_4^2  \\
m_1^2+m_2^2-p_1^2        &  2m_2^2                   &  m_2^2+m_3^2-p_2^2        &  m_2^2+m_4^2-s_{23}  \\
m_1^2+m_3^2-s_{12}  &  m_2^2+m_3^2-p_2^2        &  2m_3^2                   &  m_3^2+m_4^2-p_3^2 \\
m_1^2+m_4^2-p_4^2        &  m_2^2+m_4^2-s_{23}  &  m_3^2+m_4^2-p_3^2        &  2m_4^2
\end{array} \right)
~,
\end{equation}
%
and the matrix $Q$ in \Equation{Eq542}
%
\begin{equation}
\frac{1}{2}
\left( \begin{array}{cccc}
0  &  0                   &  0                   &  0 \\
0  &  2p_1^2              &  s_{12}+p_1^2-p_2^2  &  p_1^2+p_4^2-s_{23} \\
0  &  s_{12}+p_1^2-p_2^2  &  2s_{12}             &  s_{12}+p_4^2-p_3^2 \\
0  &  p_1^2+p_4^2-s_{23}  &  s_{12}+p_4^2-p_3^2  &  2p_4^2
\end{array} \right)
~.
\end{equation}
%

The numbering of the IR-divergent box integrals refers to the numbering in \cite{Ellis:2007qk}.
Some of the separately labeled box integrals in that publication are just particular cases of other box integrals with one or more propagator masses equal to zero.
Only the general cases are drawn below.
The graphs follow the same conventions as in \cite{Ellis:2007qk}.
Dashed lines correspond to vanishing invariants.
Explicitly mentioned invariants are non-zero and not equal to other explicitly mentioned invariants.
External lines with the same color as internal lines, and with the same label in the form of a squared mass, indicate that the squared external momentum is equal to the squared internal mass.

\paragraph{Box (16) and (15):} one soft and no collinear singularity.
$m_3^2$ may vanish.
\begin{center}
\epsfig{figure=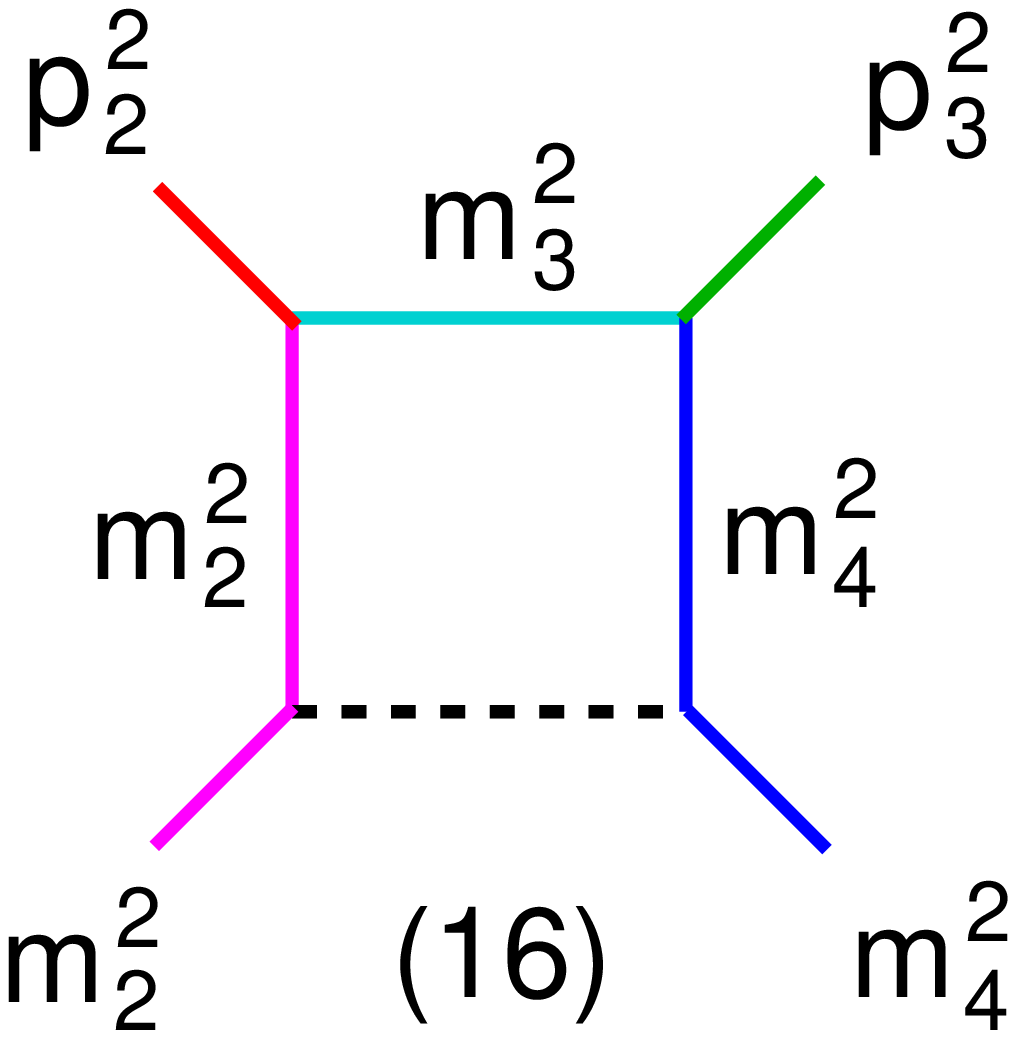,width=0.19\linewidth}
\end{center}
\begin{align}
\vc &= \frac{1}{s_{12}-m_3^2}\,\left(0\,,\,0\,,\,1\,,\,0\right)
\nonumber\\
\alpha &= \frac{1}{s_{12}-m_3^2}
\\
\vbeta &= \frac{1}{s_{12}-m_3^2}\left(0\,,\,s_{12}-p_2^2+m_2^2\,,\,2s_{12}\,,\,s_{12}-s_{3}+m_4^2\right)
~.
\nonumber\end{align}

\paragraph{Box (14):} two soft and no collinear singularities.
\begin{center}
\epsfig{figure=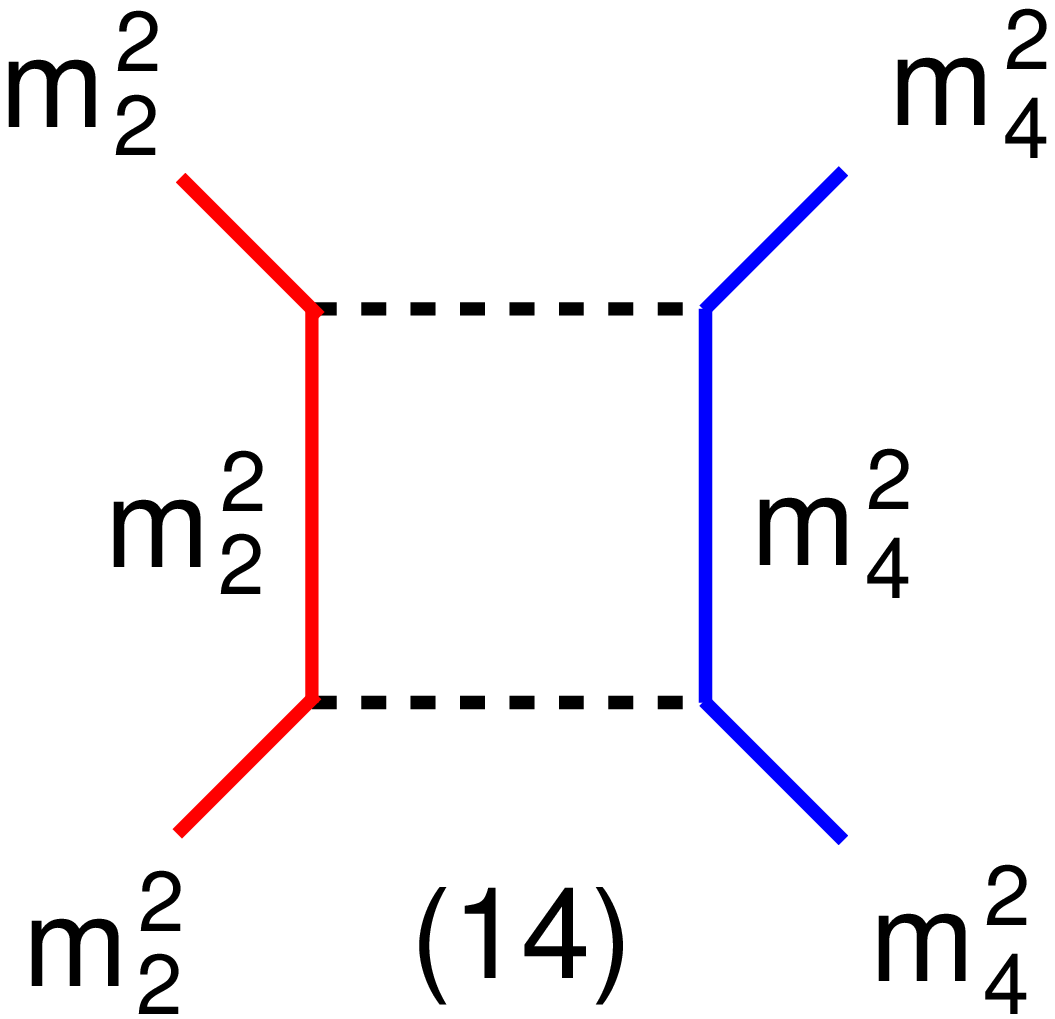,width=0.20\linewidth}
\end{center}
\begin{align}
\vc = \frac{1}{s_{12}}
       \left(1
        \,,\,0
        \,,\,1
        \,,\,0\right)
\qquad
\alpha = \frac{2}{s_{12}}
\qquad
\vbeta = \left(0
           \,,\,1
           \,,\,2
           \,,\,1\right)
~.
\end{align}

\paragraph{Box (13), (12), (11), (10), (9) and (5):} one collinear singularity, and no, one, or two soft singularities.
$m_3^2$ and $m_4^2$ may vanish in box~(13), and $m_4^2$ may vanish in box~(12).
\begin{center}
\epsfig{figure=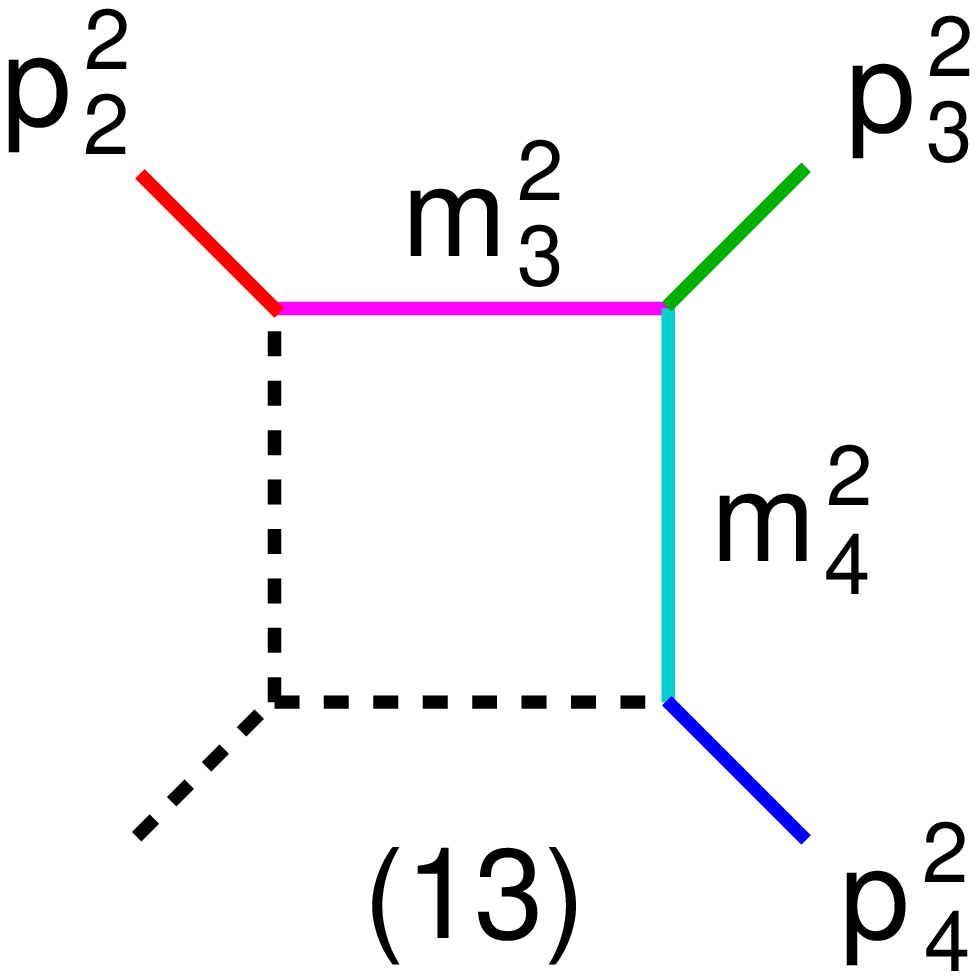,width=0.19\linewidth}\qquad
\epsfig{figure=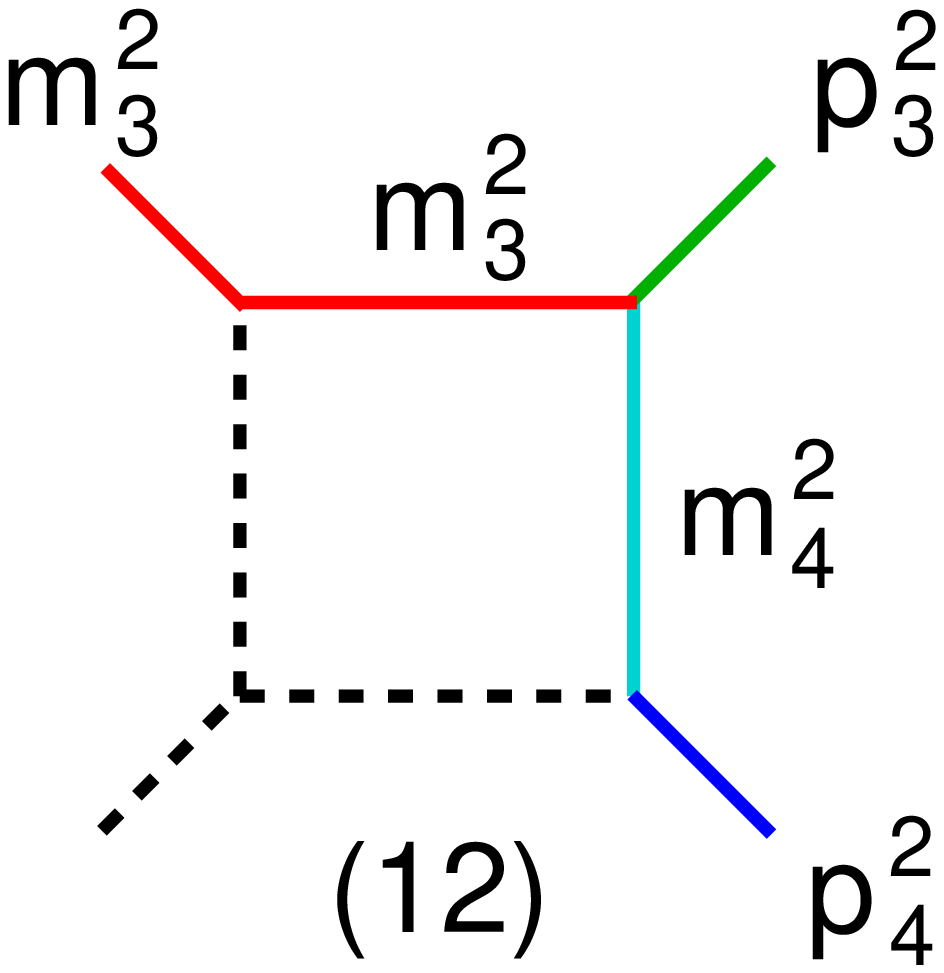,width=0.185\linewidth}\qquad
\epsfig{figure=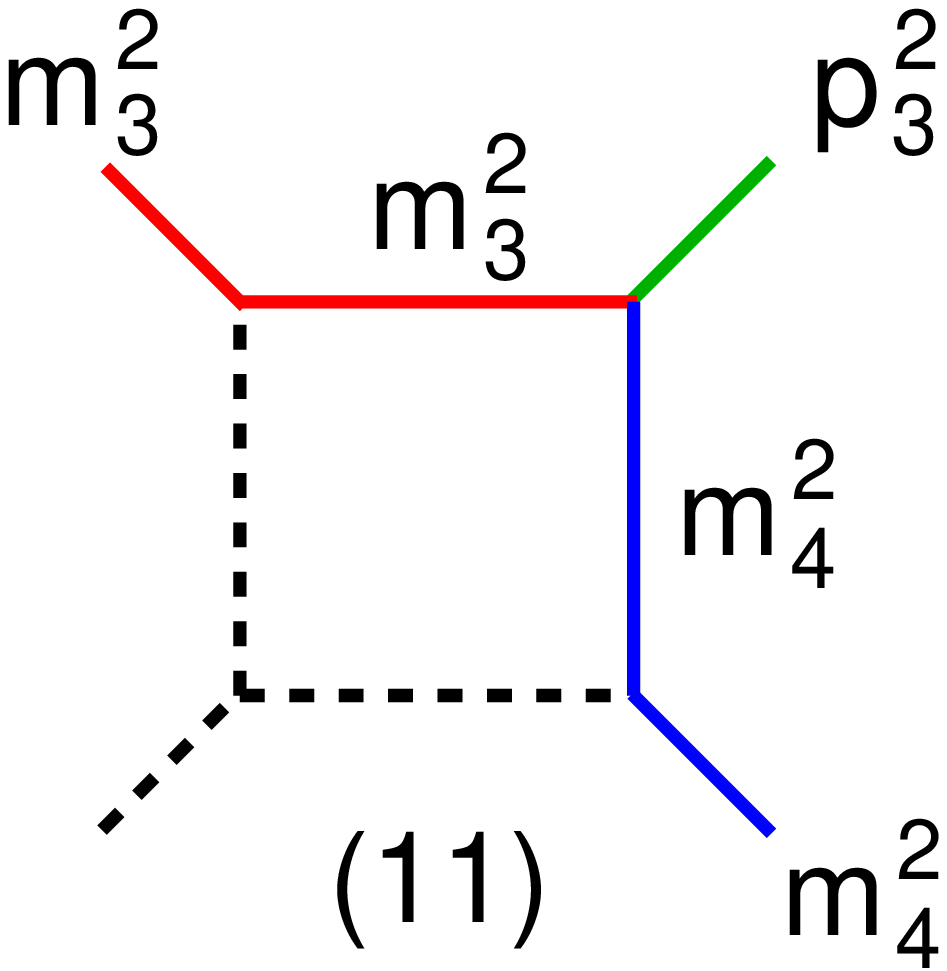,width=0.19\linewidth}
\end{center}
\begin{align}
\Delta &= (s_{12}-m_3^2)(s_{23}-m_4^2) - (p_2^2-m_3^2)(p_4^2-m_4^2)
\nonumber\\
\vc &= \frac{1}{\Delta}
       \left(0
        \,,\,0
        \,,\,s_{23}-p_4^2
        \,,\,s_{12}-p_2^2
      \right)
\nonumber\\
\alpha &= \frac{s_{12}+s_{23}-p_2^2-p_4^2}{\Delta}
\\
\vbeta &= \frac{1}{\Delta}
          \,(0
           \,,\,0
           \,,\,s_{12}(s_{12}+2s_{23}-p_4^2-p_3^2-p_2^2) + p_2^2(p_3^2-p_4^2)
\nonumber\\&\hspace{55pt}
           \,,\,p_4^2(s_{12}+s_{23}+p_3^2-p_4^2-2p_2^2) + s_{23}(s_{12}-p_3^2)
            )
~.
\nonumber\end{align}

\paragraph{Box (8), (7), (6) and (4):} two collinear and at least one soft singularity.
$m_4^2$ may vanish in box~(8).
\begin{center}
\epsfig{figure=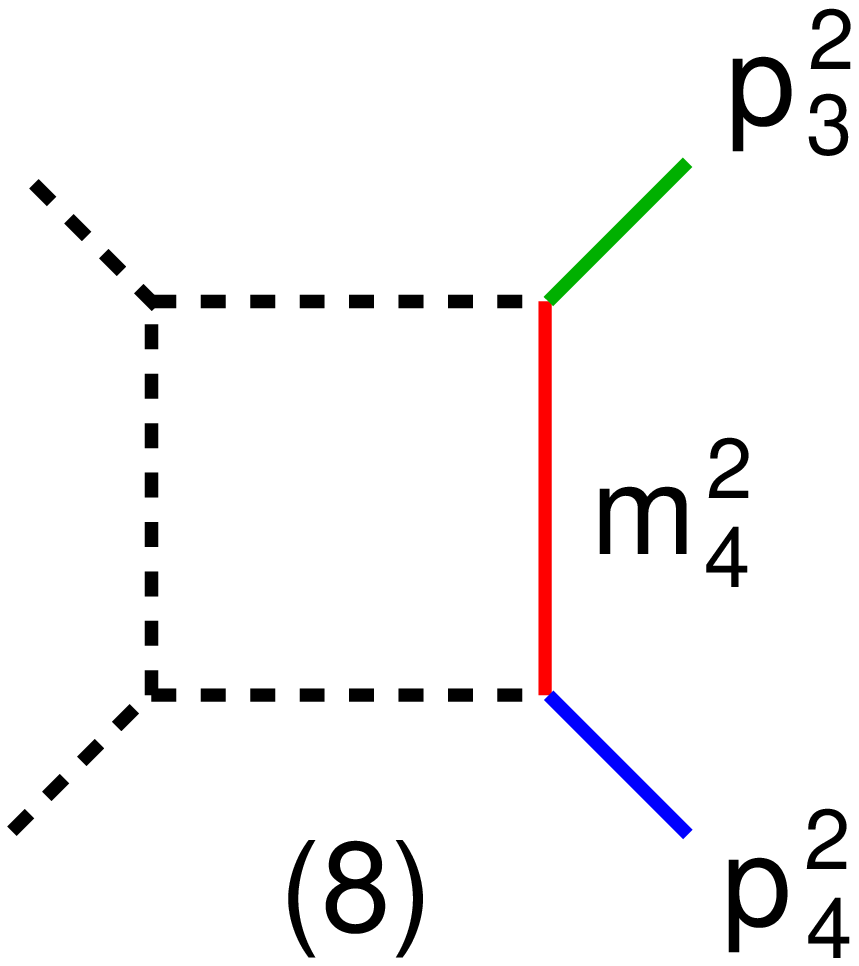,width=0.16\linewidth}\qquad
\epsfig{figure=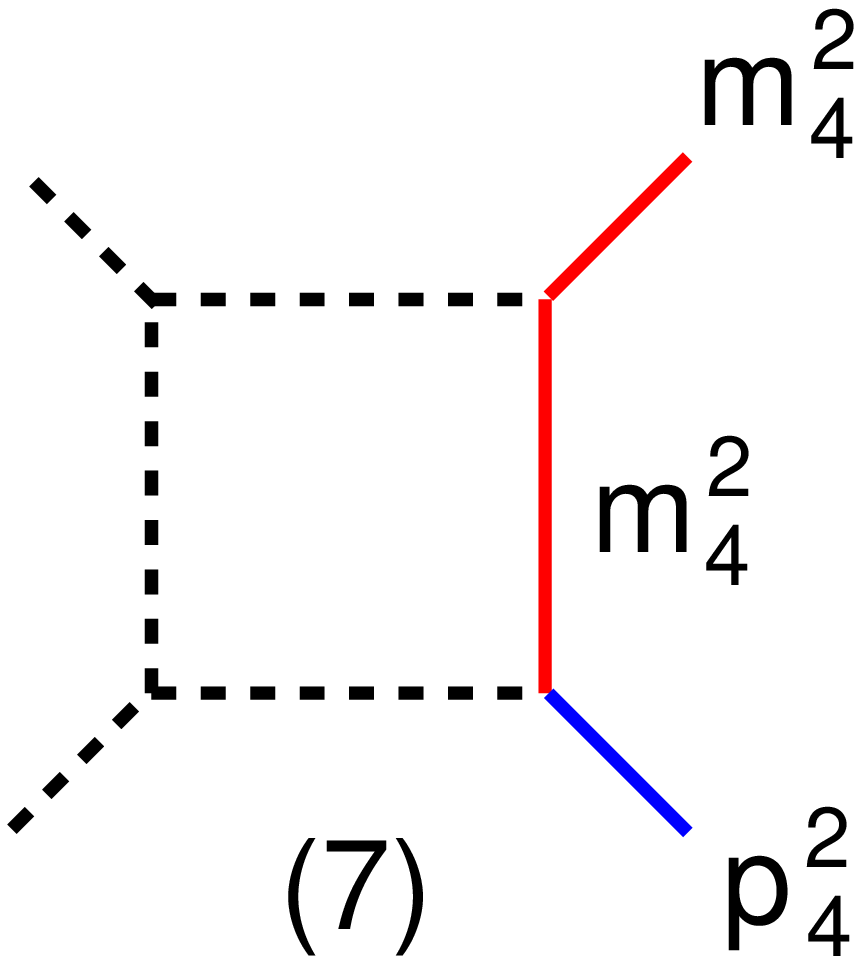,width=0.165\linewidth}\qquad
\epsfig{figure=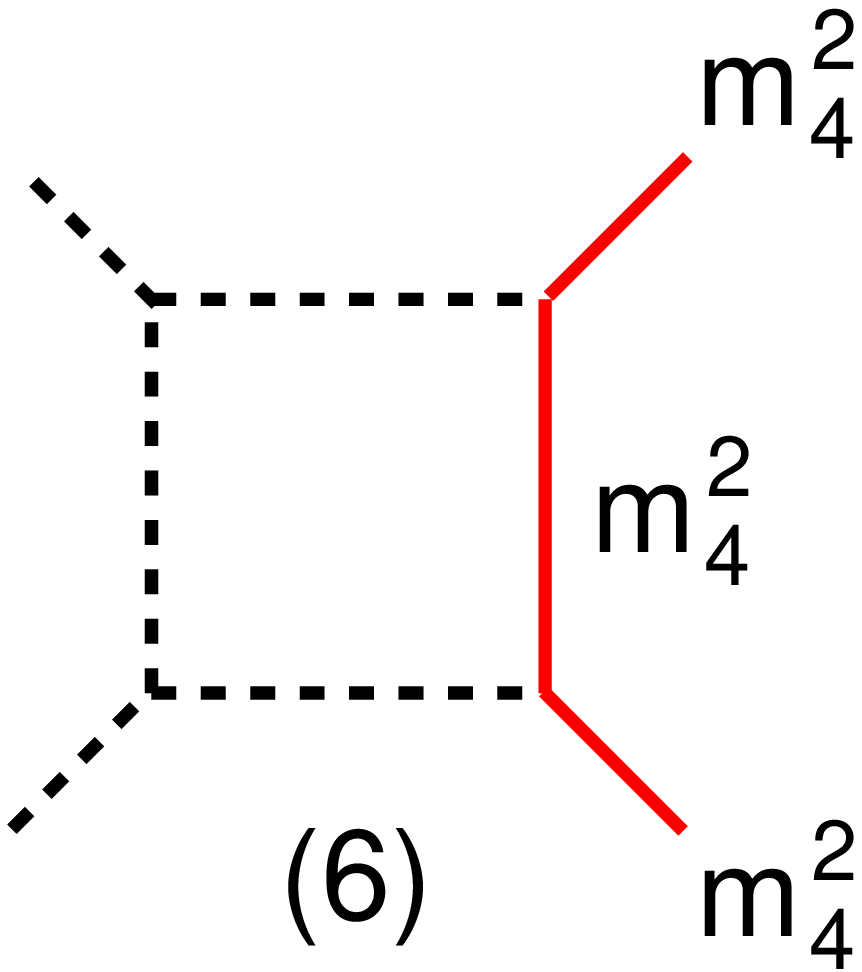,width=0.165\linewidth}
\end{center}
\begin{align}
\vc &= \frac{1}{s_{12}(s_{23}-m_4^2)}
       \left(s_{23}-p_3^2
        \,,\,0
        \,,\,s_{23}-p_4^2
        \,,\,s_{12}
      \right)
\nonumber\\
\alpha &= \frac{s_{12}+2s_{23}-p_4^2-p_3^2}{s_{12}(s_{23}-m_4^2)}
\\
\vbeta &= \frac{1}{s_{12}(s_{23}-m_4^2)}
          \,(0
           \,,\,0
           \,,\,s_{12}(s_{12}+2s_{23}-p_4^2-p_3^2)
\nonumber\\&\hspace{113pt}
           \,,\,p_4^2(s_{12}+s_{23}+p_3^2-p_4^2) + s_{23}(s_{12}-p_3^2)
            )
~.
\nonumber\end{align}

\paragraph{Box (3), (2) and (1):} more than two collinear, or two collinear and no soft singularities.
\begin{center}
\epsfig{figure=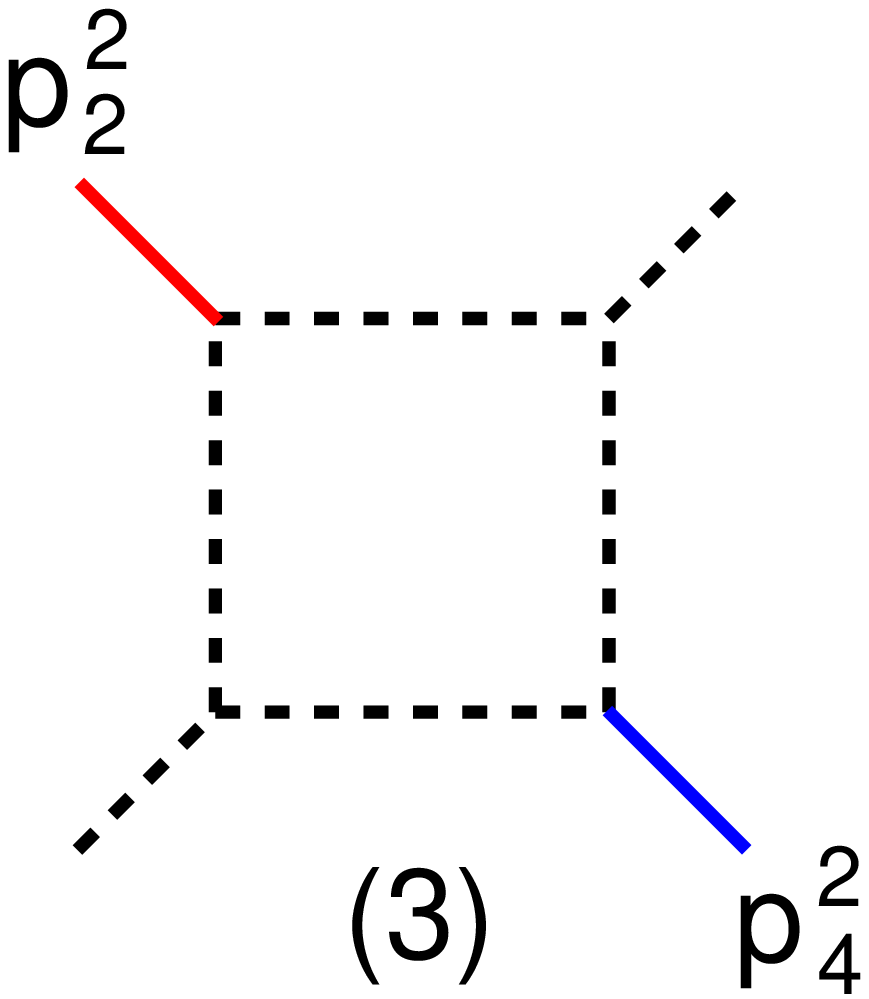,width=0.165\linewidth}\qquad
\epsfig{figure=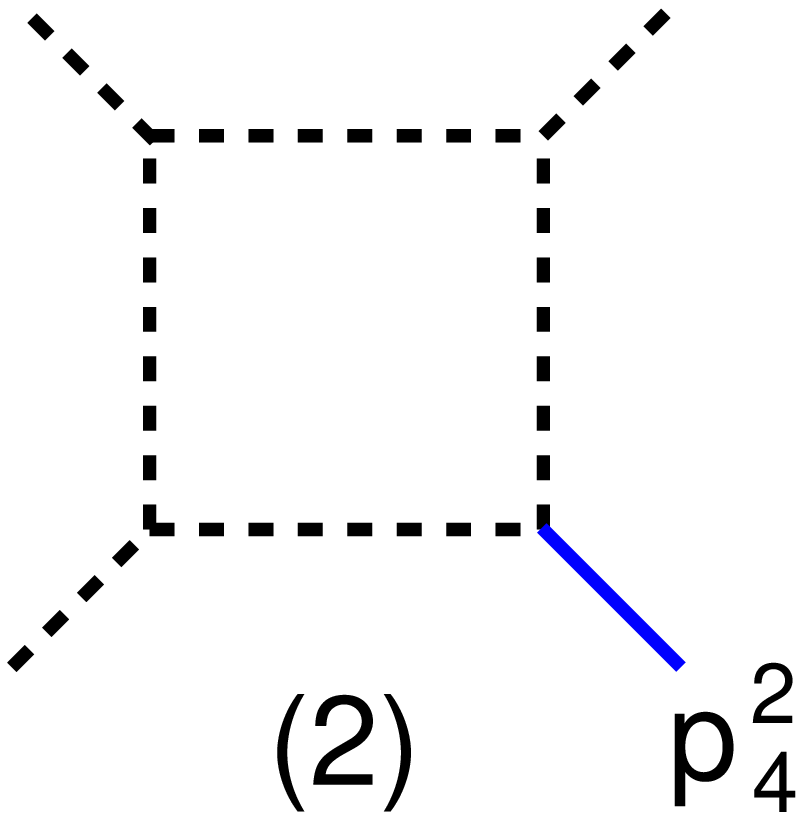,width=0.15\linewidth}\qquad
\epsfig{figure=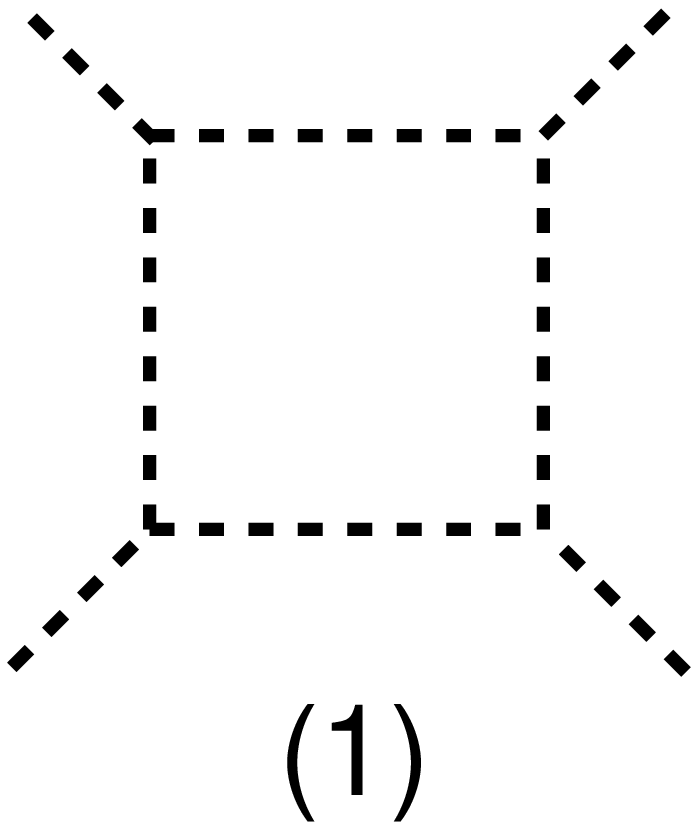,width=0.13\linewidth}
\end{center}
\begin{align}
\vc &= \frac{1}{s_{12}s_{23}-p_2^2p_4^2}
       \left(s_{23}-p_2^2
        \,,\,s_{12}-p_4^2
        \,,\,s_{23}-p_4^2
        \,,\,s_{12}-p_2^2
      \right)
\nonumber\\
\alpha &= 2\,\frac{s_{12}+s_{23}-p_2^2-p_4^2}{s_{12}s_{23}-p_2^2p_4^2}
\\
\vbeta &= \alpha
          \,(0
           \,,\,0
           \,,\,s_{12}
           \,,\,p_4^2
            )
~.
\nonumber\end{align}
%
%
\section{IR-divergent triangle integrals\label{AppTri}}
%
In this appendix the integral representations for the IR-divergent triangles are given, suitable for numerical integration.
The aim is to arrive at those with as few algebraic manipulations as possible, starting from the representation of \Equation{Eq217}.
This will be done only for triangles involving at least two mass scales (besides the extra scale $\mu$ due to the regularization), since the analytic formulas for the triangles involving only one mass scale are rather simple.
The considered triangles are depicted in \Figure{divtri}, following the labeling as in \cite{Ellis:2007qk}.
The general triangle modified Caley matrix is
%
\begin{equation}
\frac{1}{2}
\left( \begin{array}{ccc}
2m_1^2             &  m_1^2+m_2^2-p_1^2  &  m_1^2+m_3^2-p_3^2  \\
m_1^2+m_2^2-p_1^2  &  2m_2^2             &  m_2^2+m_3^2-p_2^2  \\
m_1^2+m_3^2-p_3^2  &  m_2^2+m_3^2-p_2^2  &  2m_3^2 \end{array} \right)
~.
\end{equation}
%
\myFigure{
\epsfig{figure=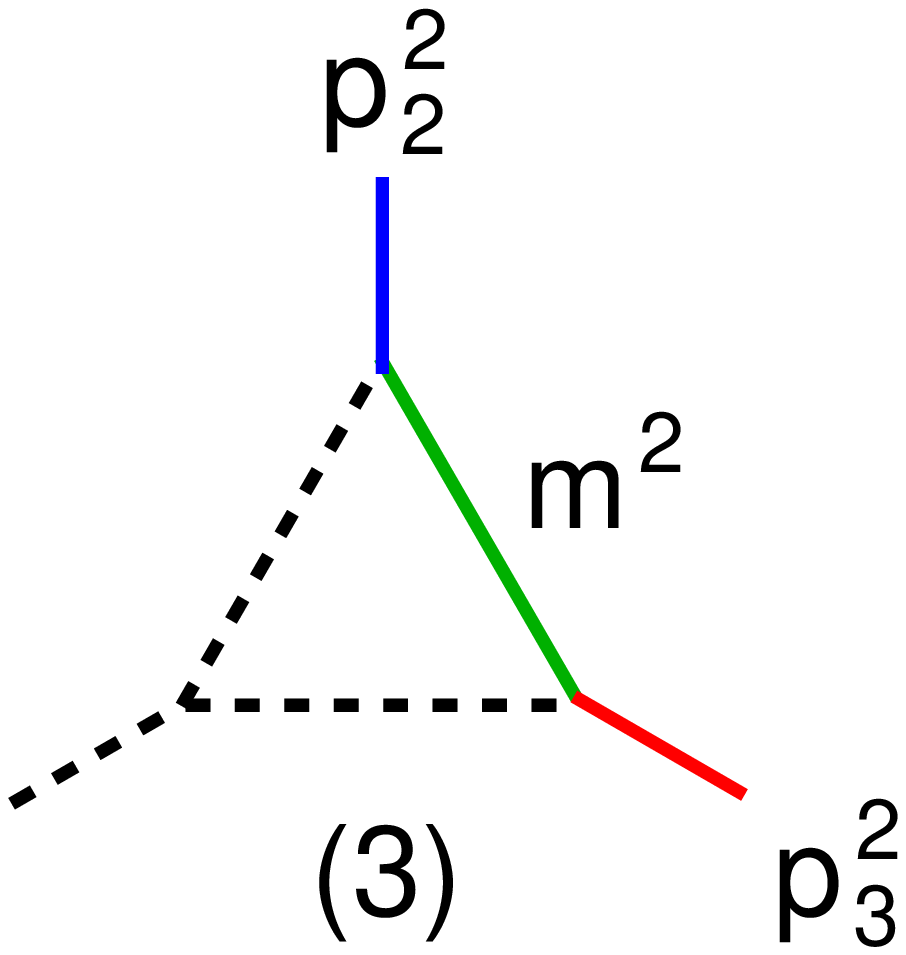,width=0.18\linewidth}\qquad
\epsfig{figure=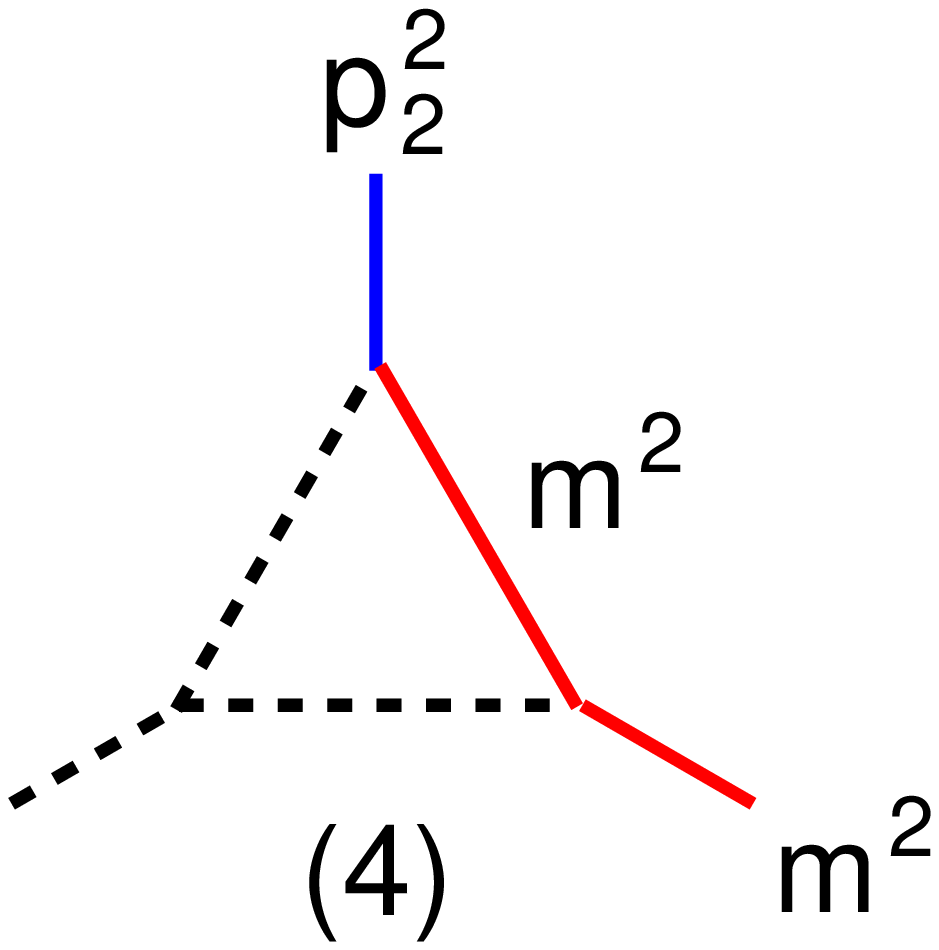,width=0.185\linewidth}\qquad
\epsfig{figure=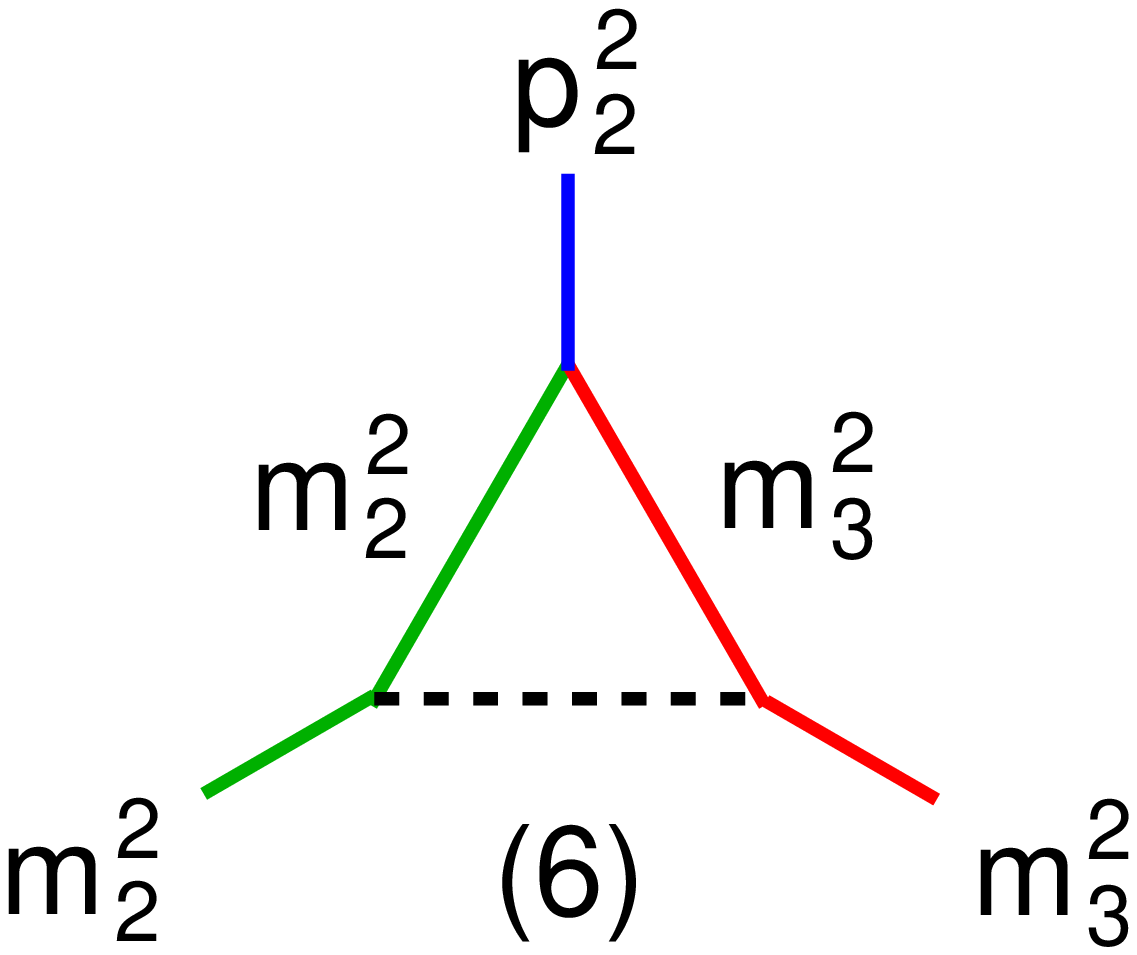,width=0.215\linewidth}
\caption{IR-divergent scalar triangles with at least two mass scales, labeled as in \cite{Ellis:2007qk}. Triangle (2) in that publication is a particular case of triangle (3) with $m^2=0$.}
\label{divtri}
}
%

\subsection*{Triangle (6)}
%
Triangle (6) has modified Caley matrix
%
\begin{equation}
\frac{1}{2}
\left( \begin{array}{ccc}
      0                 & 0                   & 0                  \\
      0                 & 2m_2^2              & m_2^2+m_3^2-p_2^2  \\
      0                 & m_2^2+m_3^2-p_2^2   & 2m_3^2 \end{array} \right)
~.
\end{equation}
%
The integral representation \Equation{Eq217} of this triangle is given by
%
\begin{equation}
\II_{3}(6)
=
\int_{0}^{\infty}dx\int_{0}^{\infty}dy
\,\frac{(1+x+y)^{-1+2\dimeps}}
       {(m_2^2y^2+(m_2^2+m_3^2-p_2^2)y+m_3^2)^{1+\dimeps}}
\end{equation}
%
The $x$-integral can be performed directly, leading to
%
\begin{equation}
\II_{3}(6)
=
\frac{-1}{2\dimeps}
\int_{0}^{\infty}dy
\,\frac{(1+y)^{2\dimeps}}
       {(m_2^2y^2+(m_2^2+m_3^2-p_2^2)y+m_3^2)^{1+\dimeps}}
~.
\end{equation}
%
The remaining integral is convergent for any value of $\dimeps$.
The integrand can be expanded into a Taylor series in $\dimeps$, which can be integrated numerically term by term.
%

\subsection*{Triangle (3)}
%
Triangle (3) has modified Caley matrix
%
\begin{equation}
\frac{1}{2}
\left( \begin{array}{ccc}
0          &  0          &  m^2-p_3^2  \\
0          &  0          &  m^2-p_2^2  \\
m^2-p_3^2  &  m^2-p_2^2  &  2m^2 \end{array} \right)
~.
\end{equation}
%
We choose to eliminate row and column 1 instead of 3 to obtain the integral representation \Equation{Eq217}, leading to
%
\begin{equation}
\II_{3}(3)
=
\int_{0}^{\infty}dx\int_{0}^{\infty}dy
\,\frac{(1+x+y)^{-1+2\dimeps}}
       {(m^2x^2+(m^2-p_2^2)xy+(m^2-p_3^2)x)^{1+\dimeps}}
~.
\end{equation}
%
Whereas the integral for triangle (6) was divergent due to the large-$x$ behavior, this integral is divergent due to the small-$x$ behavior.
Notice that a factor $x^{1+\dimeps}$ can be taken out of the denominator.
Using this fact, we apply subtraction to write
%
\begin{align}
\II_{3}(3)
&=
\int_{0}^{\infty}dx\int_{0}^{\infty}dy
\,\frac{(1+x+y)^{-1+2\dimeps}}{x^{1+\dimeps}}
\bigg[
  \frac{1}{(m^2x+(m^2-p_2^2)y+m^2-p_3^2)^{1+\dimeps}}
\nonumber\\
&\hspace{220pt}-\frac{1}{(        (m^2-p_2^2)y+m^2-p_3^2)^{1+\dimeps}}
\bigg]
\nonumber\\
&+
\frac{\pi^{1/2}\,4^{\dimeps}\,\Gamma(-\dimeps)}
     {\Gamma(\frac{1}{2}-\dimeps)}
\int_{0}^{\infty}dy
\,\frac{(1+y)^{-1+\dimeps}}{((m^2-p_2^2)y+m^2-p_3^2)^{1+\dimeps}}
~,
\label{Eq394}
\end{align}
%
where we used the identity
%
\begin{align}
\int_{0}^{\infty}dx\,\frac{(1+x+y)^{-1+2\dimeps}}{x^{1+\dimeps}}
=
\frac{\pi^{1/2}\,4^{\dimeps}\,\Gamma(-\dimeps)}
       {\Gamma(\frac{1}{2}-\dimeps)}
\,(1+y)^{-1+\dimeps}
~.
\end{align}
%
Both integrals in \Equation{Eq394} are now finite.
To obtain the $\dimeps^{-1}$-term and the $\dimeps^{0}$-term, the integrand of
the second integral may be expanded to order $\dimeps$, and $\dimeps$ may be put
to zero in the first integral.
Notice that the first, 2-dimensional, integral vanishes for $m^2=0$, \ie\ for triangle (2) in \cite{Ellis:2007qk}.
%

\subsection*{Triangle (4)}
%
The integral representation for triangle (4) becomes
%
\begin{equation}
\II_{3}(4)
=
\int_{0}^{\infty}dx\int_{0}^{\infty}dy
\,\frac{(1+x+y)^{-1+2\dimeps}}
       {(m^2x^2+(m^2-p_2^2)xy)^{1+\dimeps}}
~.
\end{equation}
%
Rescaling the variable $y\leftarrow{}xy$, we get
%
\begin{equation}
\II_{3}(4)
=
\int_{0}^{\infty}dx\int_{0}^{\infty}dy
\,\frac{(1+x+xy)^{-1+2\dimeps}}
       {x^{1+2\dimeps}(m^2+(m^2-p_2^2)y)^{1+\dimeps}}
~,
\end{equation}
%
and performing the $x$-integral, we find
%
\begin{equation}
\II_{3}(4)
=
\frac{-1}{2\dimeps}\int_{0}^{\infty}dy
\,\frac{(1+y)^{2\dimeps}}{(m^2+(m^2-p_2^2)y)^{1+\dimeps}}
~.
\end{equation}
%
Substituting $y\leftarrow1/y$ we get
%
\begin{equation}
\II_{3}(4)
=
\frac{-1}{2\dimeps}\int_{0}^{\infty}dy
\,\frac{(1+y)^{2\dimeps}}{y^{1+\dimeps}}
\,\frac{1}{(m^2y+m^2-p_2^2)^{1+\dimeps}}
~,
\end{equation}
%
which can be dealt with using subtraction
%
\begin{align}
\II_{3}(4)
=
\frac{-1}{2\dimeps}
\Bigg\{&
\int_{0}^{\infty}dy
\,\frac{(1+y)^{2\dimeps}}{y^{1+\dimeps}}
\left[ \frac{1}{(m^2y+m^2-p_2^2)^{1+\dimeps}}
      -\frac{1}{(1+y)(m^2-p_2^2)^{1+\dimeps}}\right]
\nonumber\\
&+
\frac{\pi^{1/2}\,4^{\dimeps}\,\Gamma(-\dimeps)}
     {\Gamma(\frac{1}{2}-\dimeps)}
\,\frac{1}{(m^2-p_2^2)^{1+\dimeps}}
\Bigg\}
~.
\end{align}
%
The remaining integral is finite, and can be expanded in $\dimeps$ on the integrand level.

\end{appendix}


\begin{thebibliography}{99}
%
\bibitem{Bredenstein:2008zb}
  A.~Bredenstein, A.~Denner, S.~Dittmaier and S.~Pozzorini,
  JHEP {\bf 0808} (2008) 108
  [arXiv:0807.1248 [hep-ph]].
%
\bibitem{Bredenstein:2009aj}
  A.~Bredenstein, A.~Denner, S.~Dittmaier and S.~Pozzorini,
  Phys.\ Rev.\ Lett.\  {\bf 103} (2009) 012002
  [arXiv:0905.0110 [hep-ph]].
%
\bibitem{Bredenstein:2010rs}
  A.~Bredenstein, A.~Denner, S.~Dittmaier and S.~Pozzorini,
  JHEP {\bf 1003} (2010) 021
  [arXiv:1001.4006 [hep-ph]].
%
\bibitem{Ellis:2009zw}
  R.~K.~Ellis, K.~Melnikov and G.~Zanderighi,
  JHEP {\bf 0904} (2009) 077
  [arXiv:0901.4101 [hep-ph]].
%
\bibitem{KeithEllis:2009bu}
  R.~Keith Ellis, K.~Melnikov and G.~Zanderighi,
  Phys.\ Rev.\  D {\bf 80} (2009) 094002
  [arXiv:0906.1445 [hep-ph]].
%
\bibitem{Melnikov:2009wh}
  K.~Melnikov and G.~Zanderighi,
  Phys.\ Rev.\  D {\bf 81} (2010) 074025
  [arXiv:0910.3671 [hep-ph]].
%
\bibitem{Berger:2009zg}
  C.~F.~Berger {\it et al.},
  Phys.\ Rev.\ Lett.\  {\bf 102} (2009) 222001
  [arXiv:0902.2760 [hep-ph]].
%
\bibitem{Berger:2009ep}
  C.~F.~Berger {\it et al.},
  Phys.\ Rev.\  D {\bf 80} (2009) 074036
  [arXiv:0907.1984 [hep-ph]].
%
\bibitem{Berger:2010vm}
  C.~F.~Berger {\it et al.},
  arXiv:1004.1659 [hep-ph].
%
\bibitem{Bevilacqua:2009zn}
  G.~Bevilacqua, M.~Czakon, C.~G.~Papadopoulos, R.~Pittau and M.~Worek,
  JHEP {\bf 0909} (2009) 109
  [arXiv:0907.4723 [hep-ph]].
%
\bibitem{Bevilacqua:2010ve}
  G.~Bevilacqua, M.~Czakon, C.~G.~Papadopoulos and M.~Worek,
  Phys.\ Rev.\ Lett.\  {\bf 104} (2010) 162002
  [arXiv:1002.4009 [hep-ph]].
%
\bibitem{Binoth:2009rv}
  T.~Binoth, N.~Greiner, A.~Guffanti, J.~Reuter, J.~P.~Guillet and T.~Reiter,
  Phys.\ Lett.\  B {\bf 685} (2010) 293
  [arXiv:0910.4379 [hep-ph]].
%
\bibitem{Ossola:2007ax}
  G.~Ossola, C.~G.~Papadopoulos and R.~Pittau,
  JHEP {\bf 0803} (2008) 042
  [arXiv:0711.3596 [hep-ph]].
%
\bibitem{Mastrolia:2010nb}
  P.~Mastrolia, G.~Ossola, T.~Reiter and F.~Tramontano,
  arXiv:1006.0710 [hep-ph].


%
\bibitem{Ellis:2007qk}
  R.~K.~Ellis and G.~Zanderighi,
  JHEP {\bf 0802} (2008) 002
  [arXiv:0712.1851 [hep-ph]].
%
\bibitem{vanHameren:2009dr}
  A.~van Hameren, C.~G.~Papadopoulos and R.~Pittau,
  JHEP {\bf 0909}, 106 (2009)
  [arXiv:0903.4665 [hep-ph]].
%
\bibitem{Hahn:1998yk}
  T.~Hahn and M.~Perez-Victoria,
  Comput.\ Phys.\ Commun.\  {\bf 118}, 153 (1999)
  [arXiv:hep-ph/9807565].
%
\bibitem{vanOldenborgh:1989wn}
  G.~J.~van Oldenborgh and J.~A.~M.~Vermaseren,
  Z.\ Phys.\  C {\bf 46} (1990) 425.
%
\bibitem{Denner:2005fg}
  A.~Denner, S.~Dittmaier, M.~Roth and L.~H.~Wieders,
  Nucl.\ Phys.\  B {\bf 724} (2005) 247
  [arXiv:hep-ph/0505042].
%
\bibitem{Nhung:2009pm}
  D.~T.~Nhung and L.~D.~Ninh,
  Comput.\ Phys.\ Commun.\  {\bf 180} (2009) 2258
  [arXiv:0902.0325 [hep-ph]].
%
\bibitem{'tHooft:1978xw}
  G.~'t Hooft and M.~J.~G.~Veltman,
  Nucl.\ Phys.\  B {\bf 153} (1979) 365.
%
\bibitem{Denner:2010tr}
  A.~Denner and S.~Dittmaier,
  arXiv:1005.2076 [hep-ph].
%
%
\bibitem{colilib}
  W.~Beenakker and A.~Denner,
  Nucl.\ Phys.\  B {\bf 338} (1990) 349.
  A.~Denner, U.~Nierste and R.~Scharf,
  Nucl.\ Phys.\  B {\bf 367}, 637 (1991).
  A.~Denner and S.~Dittmaier,
  Nucl.\ Phys.\  B {\bf 658} (2003) 175
  [arXiv:hep-ph/0212259].
  A.~Denner and S.~Dittmaier,
  Nucl.\ Phys.\  B {\bf 734} (2006) 62
  [arXiv:hep-ph/0509141].
%
%
\bibitem{Hahn:2004fe}
  T.~Hahn,
  Comput.\ Phys.\ Commun.\  {\bf 168} (2005) 78
  [arXiv:hep-ph/0404043].
%
\bibitem{quadpack}
R.~Piessens, E.~de~Doncker, C.~\"Uberhuber, D.~Kahaner,
Springer-Verlag, 1983.
%
%
\bibitem{Passarino:2001wv}
  G.~Passarino,
  Nucl.\ Phys.\  B {\bf 619} (2001) 257
  [arXiv:hep-ph/0108252].
%
\bibitem{Ferroglia:2002mz}
  A.~Ferroglia, M.~Passera, G.~Passarino and S.~Uccirati,
  Nucl.\ Phys.\  B {\bf 650}, 162 (2003)
  [arXiv:hep-ph/0209219].
%
\bibitem{Nagy:2003qn}
  Z.~Nagy and D.~E.~Soper,
  JHEP {\bf 0309} (2003) 055
  [arXiv:hep-ph/0308127].
%
\bibitem{Nagy:2006xy}
  Z.~Nagy and D.~E.~Soper,
  Phys.\ Rev.\  D {\bf 74} (2006) 093006
  [arXiv:hep-ph/0610028].
%
\bibitem{Gong:2008ww}
  W.~Gong, Z.~Nagy and D.~E.~Soper,
  Phys.\ Rev.\  D {\bf 79} (2009) 033005
  [arXiv:0812.3686 [hep-ph]].
%
\bibitem{deDoncker:2004fb}
  E.~de Doncker, Y.~Shimizu, J.~Fujimoto and F.~Yuasa,
  Comput.\ Phys.\ Commun.\  {\bf 159} (2004) 145.

\bibitem{Binoth:2005ff}
  T.~Binoth, J.~P.~Guillet, G.~Heinrich, E.~Pilon and C.~Schubert,
  JHEP {\bf 0510} (2005) 015
  [arXiv:hep-ph/0504267].
%
\bibitem{Kurihara:2005ja}
  Y.~Kurihara and T.~Kaneko,
  Comput.\ Phys.\ Commun.\  {\bf 174} (2006) 530
  [arXiv:hep-ph/0503003].
%
\cite{Anastasiou:2007qb}
\bibitem{Anastasiou:2007qb}
  C.~Anastasiou, S.~Beerli and A.~Daleo,
  JHEP {\bf 0705} (2007) 071
  [arXiv:hep-ph/0703282].
%
\bibitem{Dittmaier:2003bc}
  S.~Dittmaier,
  Nucl.\ Phys.\  B {\bf 675} (2003) 447
  [arXiv:hep-ph/0308246].
%
\end{thebibliography}
\end{document}